\documentclass[aps,prb,twocolumn,superscriptaddress,showpacs]{revtex4-1}
\usepackage{graphicx}
\usepackage{color}

\begin{document}

\title{Transport, Thermal, and Magnetic Properties of the Narrow-Gap Semiconductor CrSb$_2$}

\author{Brian C. Sales}
\affiliation{Materials Science and Technology Division, Oak Ridge National Laboratory, Oak Ridge, Tennessee 37831}
\author{Andrew F. May}
\affiliation{Materials Science and Technology Division, Oak Ridge National Laboratory, Oak Ridge, Tennessee 37831}
\author{Michael A. McGuire}
\affiliation{Materials Science and Technology Division, Oak Ridge National Laboratory, Oak Ridge, Tennessee 37831}
\author{Matthew B. Stone}
\affiliation{Neutron Sciences Directorate, Quantum Condensed Matter Division, Oak Ridge National Laboratory, Oak Ridge, Tennessee 37831}
\author{David J. Singh}
\affiliation{Materials Science and Technology Division, Oak Ridge National Laboratory, Oak Ridge, Tennessee 37831}
\author{David Mandrus}
\affiliation{Materials Science and Technology Division, Oak Ridge National Laboratory, Oak Ridge, Tennessee 37831}
\affiliation{Department of Materials Science and Engineering, University of Tennessee, Knoxville, Tennessee 37996}

\begin{abstract}
Resistivity, Hall effect, Seebeck coefficient, thermal conductivity, heat capacity, and magnetic susceptibility data are reported for CrSb$_2$ single crystals. In spite of some unusual features in electrical transport and Hall measurements below 100\,K, only one phase transition is found in the temperature range from 2 to 750\,K corresponding to long-range antiferromagnetic order below T$_{\mathrm{N}}$ $\approx$ 273\,K. Many of the low temperature properties can be explained by the thermal depopulation of carriers from the conduction band into a low mobility band located approximately 16 meV below the conduction band edge, as deduced from the Hall effect data. In analogy with what occurs in Ge, the low mobility band is likely an impurity band. The Seebeck coefficient, S, is large and negative for temperatures from 2 to 300\,K ranging from $\approx$ -70\,$\mu$V/K at 300\,K to -4500\,$\mu$V/K at 18\,K. A large maximum in $|$S$|$ at 18\,K is likely due to phonon drag with the abrupt drop in $|$S$|$ below 18\,K due to the thermal depopulation of the high mobility conduction band.  The large thermal conductivity between 10 and 20\,K ($\approx$ 350\,W/m-K) is consistent with this interpretation, as are detailed calculations of the Seebeck coefficient made using the complete calculated electronic structure. These data are compared to data reported for FeSb$_2$, which crystallizes in the same marcasite structure, and FeSi, another unusual narrow-gap semiconductor.
\end{abstract}


\maketitle

\section{Introduction}

The compounds FeSi, FeSb$_2$ and CrSb$_2$ are narrow gap semiconductors, Eg $\approx$ 0.1\,eV, with dominant 3d-character of the electronic states near the conduction and valence band edges and interesting magnetic and transport properties\cite{Jaccarino_1976,Mandrus_1995_FeSi,Sales_1994_FeIrSi,Wolfe_1965,DeGiorgi_1994,Onose_Tokura_2005, Bentien_1994,Petrovic_2003,Takahashi_Terasaki_2011,Sun_2011,Holseth_1970,Alles_1978,Hu_2007,Delaire_2011,Sales_2011,Tomcsak_2012,Takahashi_Sato_2011,Wang_2012}. Both CrSb$_2$ and FeSb$_2$  crystallize in the orthorhombic marcasite structure ($Pnnm$ space group), while FeSi crystallizes in the cubic B20 structure ($P$2$_1$3 space group). At first glance, the magnetic susceptibility of all three compounds is similar with a broad maximum in the susceptibility at 550\,K, 500\,K, and 400\,K for CrSb$_2$, FeSi, and FeSb$_2$ respectively. The values of the susceptibility maxima are also similar:  $\chi_{\mathrm{max}}$ = 13.8$\times$10$^{-4}$\,cm$^3$/mole-CrSb$_2$, 8.5$\times$10$^{-4}$\,cm$^3$/mole-FeSi, and $\approx$4.5$\times$10$^{-4}$\,cm$^3$/mole-FeSb$_2$.

\begin{figure}[!ht]
\includegraphics [width=2.5in] {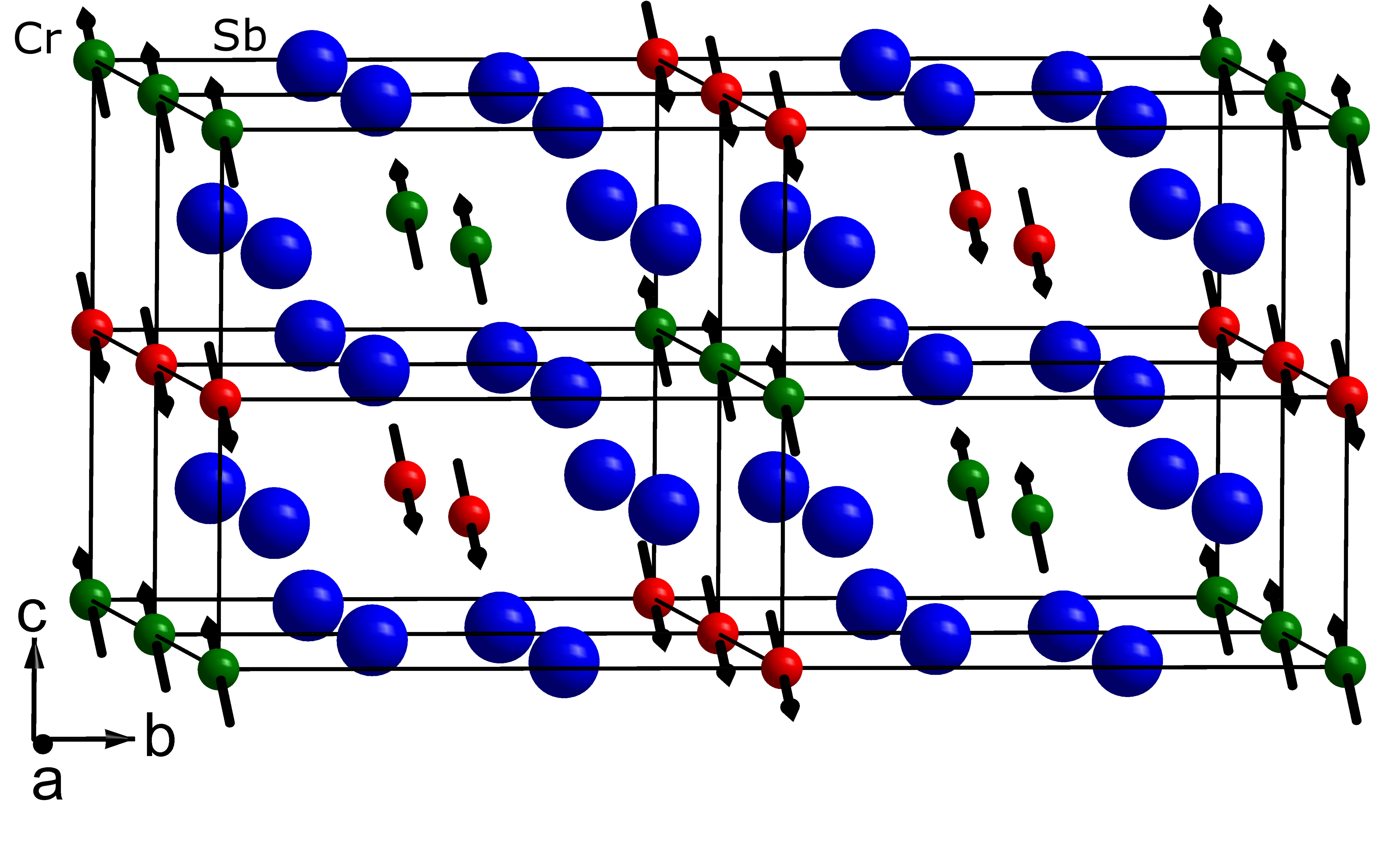}
\includegraphics [width=2.5in] {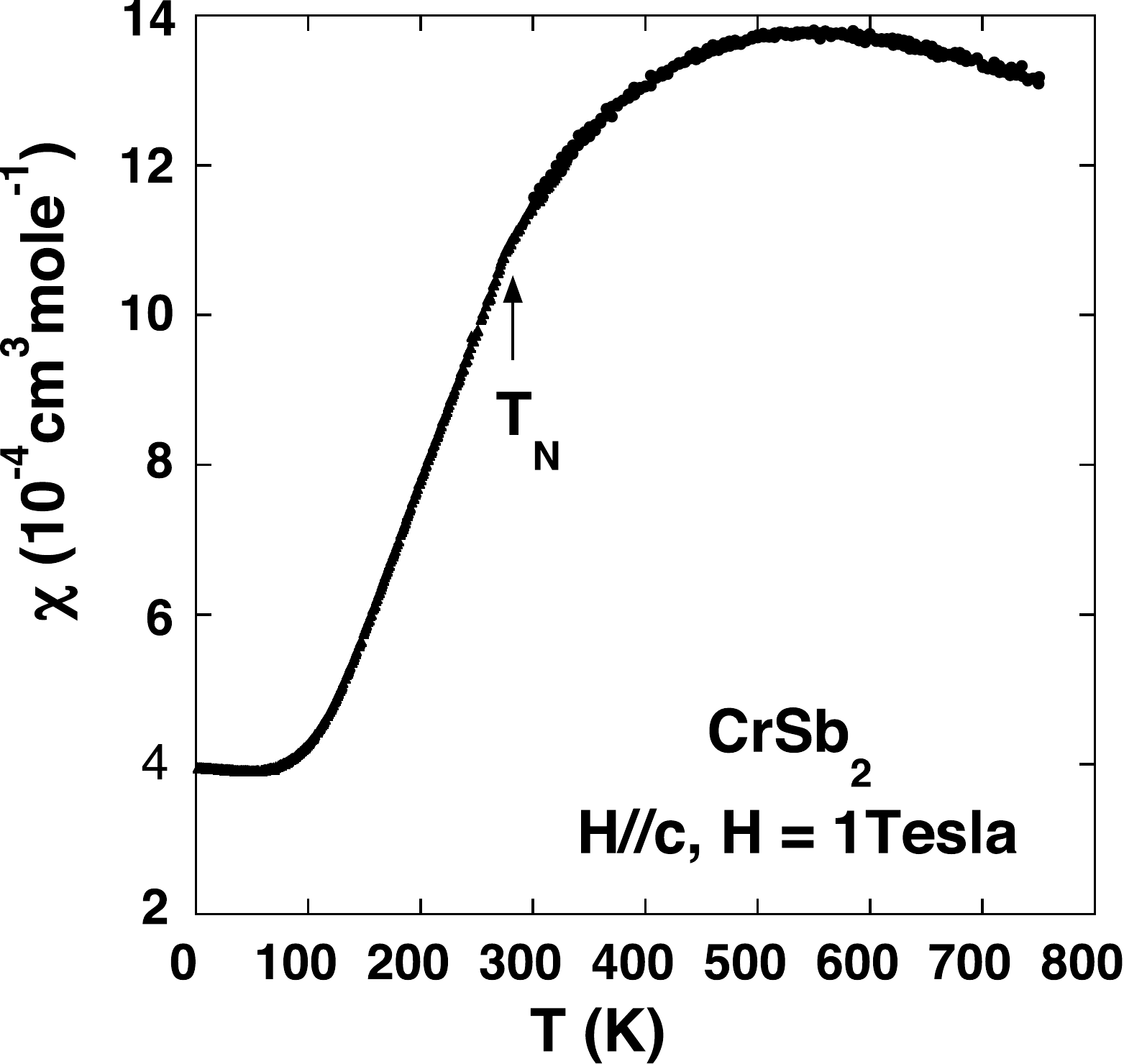}
\caption {(color online).(top) Nuclear and magnetic structure of CrSb$_2$. Red and green spheres are Cr sites, where both the vector and color represent the spin orientation; blue spheres are the Sb sites and a total of one magnetic unit cell is shown while crystallographic unit cells are outlined in thin black lines. The Cr spins are aligned approximately along [101].  (bottom) Magnetic susceptibility vs. temperature for a CrSb$_2$ crystal with a magnetic field of 1 Tesla applied along the $c$ axis. The Neel temperature is noted in the figure.}
\label{sus_structure}
\end{figure}

Unlike FeSb$_2$ and FeSi, however, CrSb$_2$ exhibits antiferromagnetic order (T$_{\mathrm{N}}\approx$\,273\,K).  The observed magnetic structure is shown in Fig.\ref{sus_structure}a, where each Cr has a magnetic moment of 1.9$\mu_{\mathrm{B}}$.\cite{Holseth_1970,Stone_2012}  The susceptibility data for an oriented crystal of CrSb$_2$ from 2 to 750\,K are shown in Fig.\ref{sus_structure}b.  The weak anomalies in magnetic susceptibility and resistivity near T$_{\mathrm{N}}$ are due to the quasi-1D nature of the magnetism, as was proved recently using inelastic neutron scattering measurements.\cite{Stone_2012}   The nearest-neighbor exchange constants were found to be \textbf{J}$_{\mathrm{\textbf{c}}} \approx$36\,meV, while \textbf{J}$_{\mathrm{\textbf{a}}} \approx$-1.3\,meV (ferromagnetic), \textbf{J}$_{\mathrm{\textbf{b}}} \approx$1.5\,meV and \textbf{J}$_{\mathrm{\textbf{111}}} \approx$0.\cite{Stone_2012} Most of the entropy associated with the S=1 Cr spins is removed above T$_{\mathrm{N}}$ due to short range order primarily along the $c$ axis.\cite{Stone_2012}

In this article, we report low temperature electrical and thermal transport data, heat capacity and magnetic susceptibility measurements from oriented CrSb$_2$ single crystals.  In the temperature range 15$-$20\,K, well below the magnetic ordering temperature, we find an unexpected minimum in the carrier concentration inferred from Hall effect data, and a very large Seebeck coefficient (reaching -4500 $\mu$V/K). It will be shown that the resistivity and Hall effect data for CrSb$_2$ can be explained by a low mobility impurity band located $\approx$16\,meV below the conduction band edge, and that phonon-drag is likely to be responsible for the large magnitude of the Seebeck coefficient.

\section{Experimental Methods}
Single crystals of CrSb$_2$ are grown out of an Sb flux. High purity Sb shot (99.9999\% Alfa Aesar) and Cr powder (99.99\% Alfa Aesar) in the molar ratio 94:6 are loaded into a 10\,cc alumina crucible, and then vacuum-sealed in a quartz ampoule. The mixture is heated to 1000$^{\circ}$C over 6\,h, held for 36\,h and then cooled slowly (2$^{\circ}$C/h) to 640$^{\circ}$C. The quartz ampoule is quickly removed from the furnace and placed in a centrifuge where the excess Sb is removed. The crystals grown via this method are sometimes $\approx$ 1$\times$1$\times$1\,cm$^3$, which are large enough for thermal transport and electrical measurements using the Thermal Transport Option (TTO) on the Physical Property Measurement System (PPMS) from Quantum Design. Powder x-ray diffraction and energy dispersive x-ray analysis confirmed that the crystals were single phase with the orthorhombic marcasite structure reported previously \cite{Holseth_1970,Stone_2012}. Oriented single crystals, with typical dimensions of 9$\times$1.3$\times$1.3\,mm$^3$, are cut from the as-grown crystals using a low-speed diamond saw (different crystals were utilized to obtain transport data along the different crystallographic directions, and properties were found to be consistent between batches). Crystals with the long dimension along [001] ($c$ axis) and along [110] (perpendicular to $c$ axis) are investigated.  Leads for electrical and thermal measurements are attached with conducting epoxy and silver paste so that the contact resistance is below 5\,$\Omega$. For the TTO measurements the typical contact area for the four electrical and thermal leads is about 1.2 $\times$ 1.2\,mm$^2$. Varying the applied temperature difference during TTO measurements from about 20 to 200 mK resulted in less than 5\% variation in the derived value of the Seebeck coefficient and thermal conductivity. Additional Hall effect, resistivity, and magnetoresistance measurements are made on thinned ($\approx$0.5\,mm thick) oriented rectangular crystal plates. For each of these measurements electrical contact is made with 0.025\,mm diameter Pt wires and contact areas of about 0.1 $\times$ 0.1\,mm$^2$. Standard four lead geometries are used for all electrical measurements with the magnetic field, \textbf{H}, applied perpendicular to the current direction. For the normal resistivity measurements the current direction is reversed so that any thermal voltages can be eliminated. For the Hall measurements, the applied magnetic field is swept from positive to negative values so that the contribution of the normal resistivity ($\rho_{xx}$) due to any slight error in the position of the Hall leads could be eliminated. The Hall resistivity, $\rho_{xy}$, is then given by [$\rho_{xy}$(H)-$\rho_{xy}$(-H)]/2.\cite{Sales_2006}  Thermal conductivity, Seebeck coefficient and electrical resistivity measurements are made using the TTO option, and additional resistivity and Hall effect measurements are made using the PPMS resistivity option. Heat capacity and magnetic susceptibility measurements are made with the PPMS heat capacity option, and a SQUID magnetometer (MPMS) from Quantum Design.

\section{Results and Discussion}

\begin{figure}[!ht]
\includegraphics [width=3in] {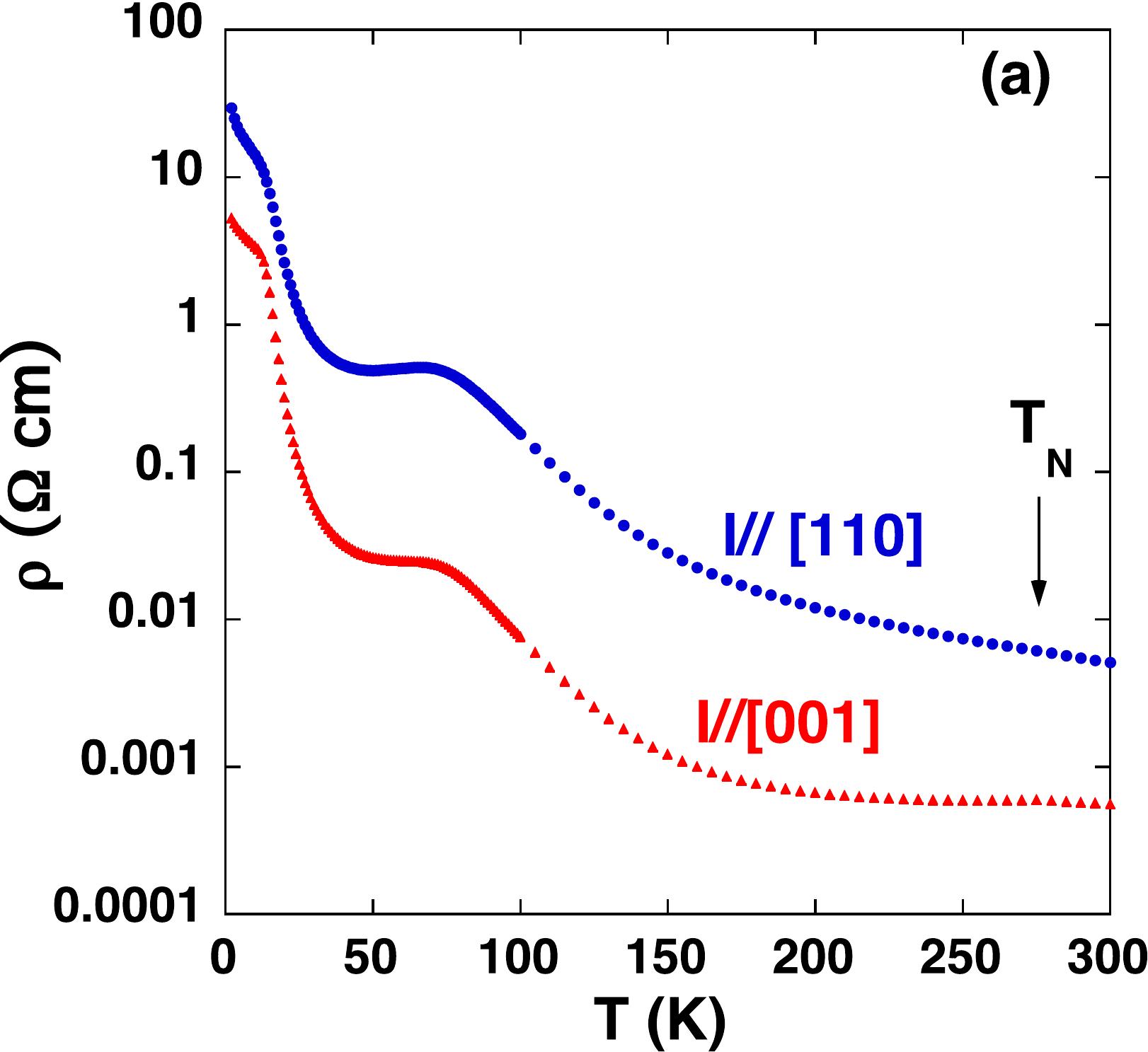}
\includegraphics [width=2.85in] {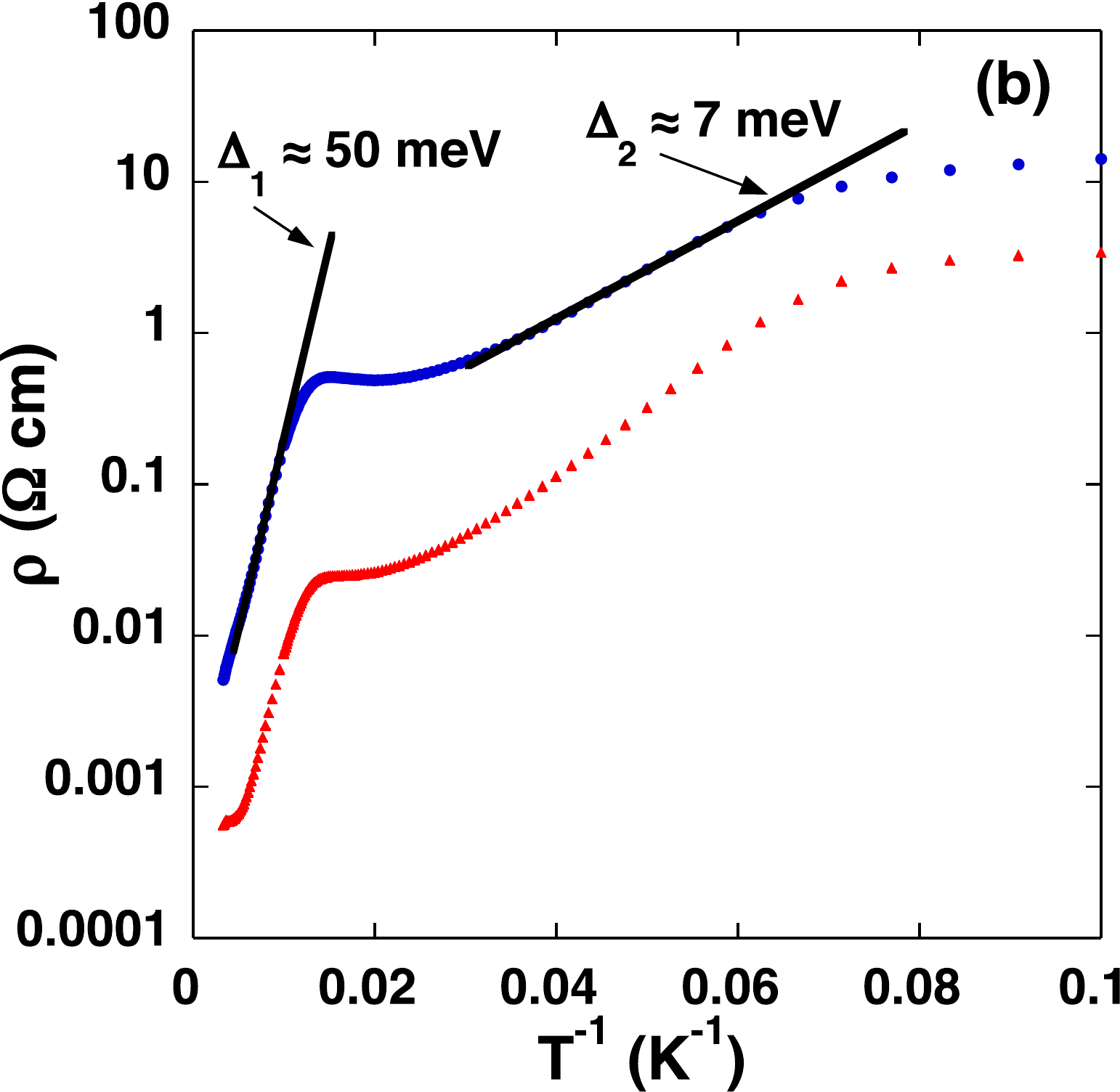}
\caption {(color online).(a) Resistivity versus temperature for CrSb$_2$ crystals with current along [001] and [110] is consistent with lightly-doped semiconductor behavior. (b) log $\rho$  versus inverse temperature. Different regions of activated transport are noted in the figure.}
\label{resist}
\end{figure}

The resistivity data from CrSb$_2$ with the current along the $c$ axis and along [110] are shown in Fig.\,\ref{resist}. The resistivity along the $c$ axis is about one order of magnitude lower than the resistivity along [110]. Overall, however, the temperature dependence of the resistivity is similar for the two directions. These data are similar to that reported by Hu et al. both in magnitude and shape.\cite{Hu_2007}  There is a small feature in the resistivity near T$_{\mathrm{N}}$$\approx$273\,K (not shown).

Activated behavior is observed for transport along both directions at temperatures below 200\,K. The resistivity data can be described by $\rho \propto$ exp($\Delta$/k$_{\mathrm{B}}$T) over limited temperature ranges: $\Delta_1 \approx$ 50\,meV for  100\,K $<$T$<$300\,K; $\Delta_2 \approx$ 7-9\,meV (depending on crystal orientation) for 16\,K$<$T$<$33\,K, and $\Delta_3 <$0.1\, meV for T$<$10\,K (not shown). The high temperature gap (2$\Delta_1$ $\approx$ 100\,meV) is close to the intrinsic value of 140\,meV estimated from resistivity measurements on polycrystalline samples or single crystals\cite{Hu_2007,Harada_2004} above room temperature.

The plateau in the resistivity curves near 50-80\,K suggests a possible phase transition in this temperature region. There is no evidence, however, of a phase transition in x-ray, neutron and heat capacity data.\cite{Alles_1978,Stone_2012}  The behavior between 50-80\,K is likely related to the decrease in carrier mobility with increasing temperature when phonon scattering dominates the carrier relaxation time. This scattering effect is generally observable when the carrier density in the conduction band is relatively constant, which is possible in this temperature range considering the magnitude of the high-temperature band gap and the analysis of the Hall data presented below.  Thus, the temperature dependence of the resistivity is consistent with a lightly-doped, narrow gap semiconductor: impurity band conduction dominates at low T, extrinsic carriers in the conduction band dominate at moderate T, and intrinsic conduction dominates at high T.\cite{Neamen_1992} This type of plateau in resistivity data is also observed in doped Si and Ge crystals when the carrier concentration is in the range of 10$^{17}$\,carriers/cm$^3$,\cite{Yamanouchi_1967,Fritzsche_1955,Swartz_1960} as well as in FeSi and FeSb$_2$ crystals\cite{Mandrus_1995_FeSi,Sales_1994_FeIrSi,Bentien_1994,Petrovic_2003,Hu_2007,Takahashi_Sato_2011} and polycrystalline CrSb$_{2}$.\cite{Li_2008,Li_2009}

\begin{figure*}[t]
\includegraphics [height=2.1in] {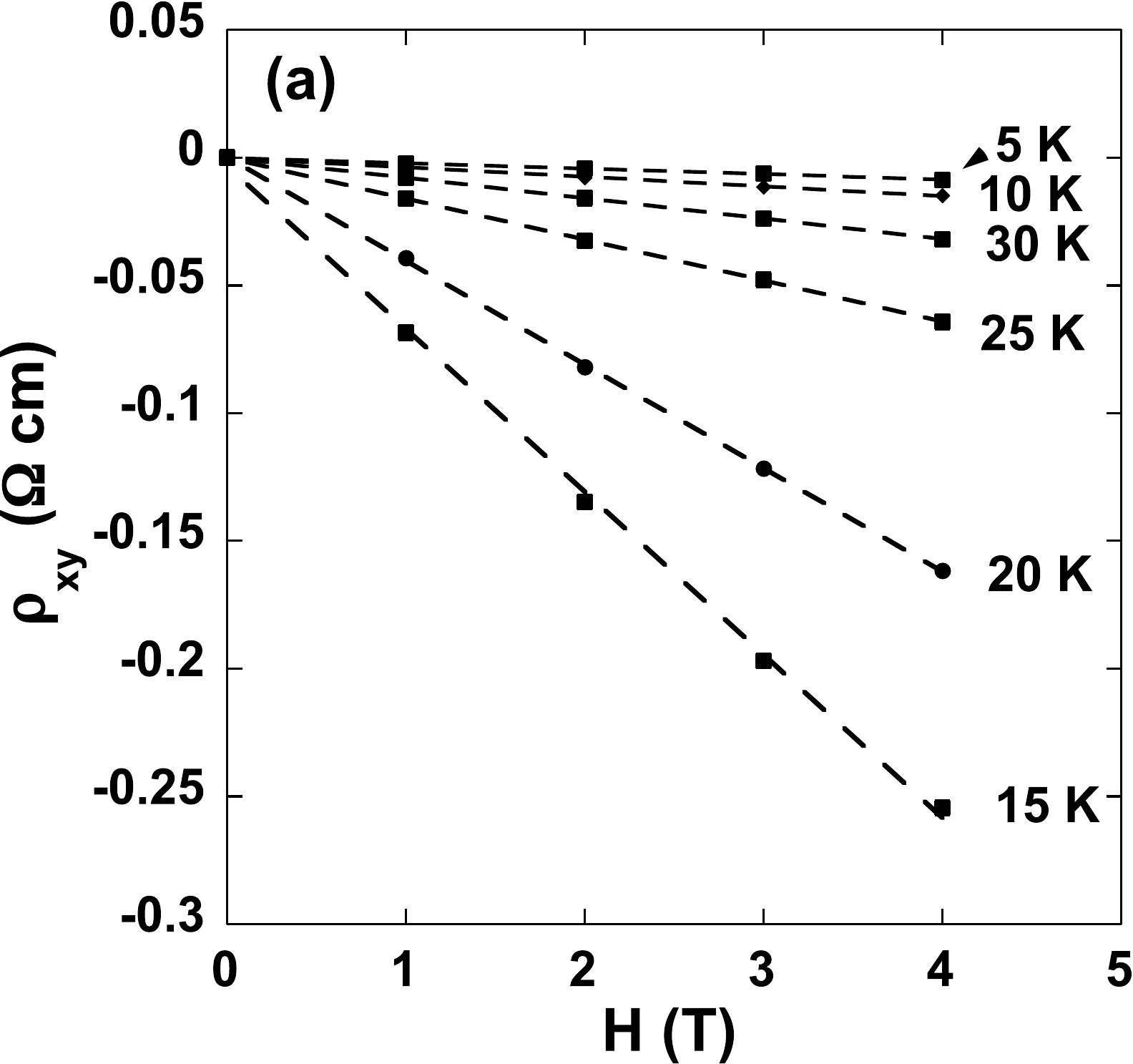}
\includegraphics [height=2.1in] {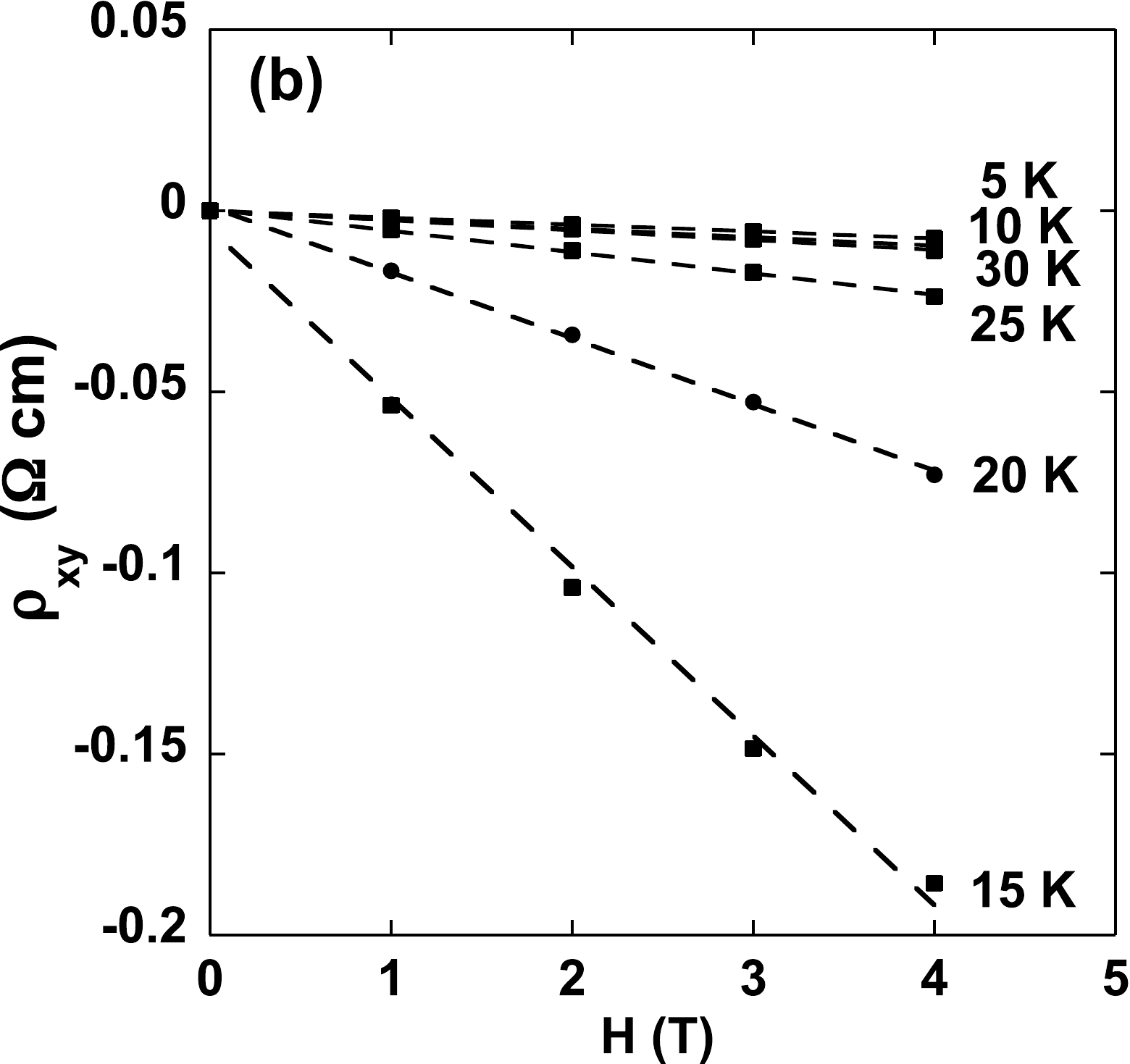}
\includegraphics [height=2.14in] {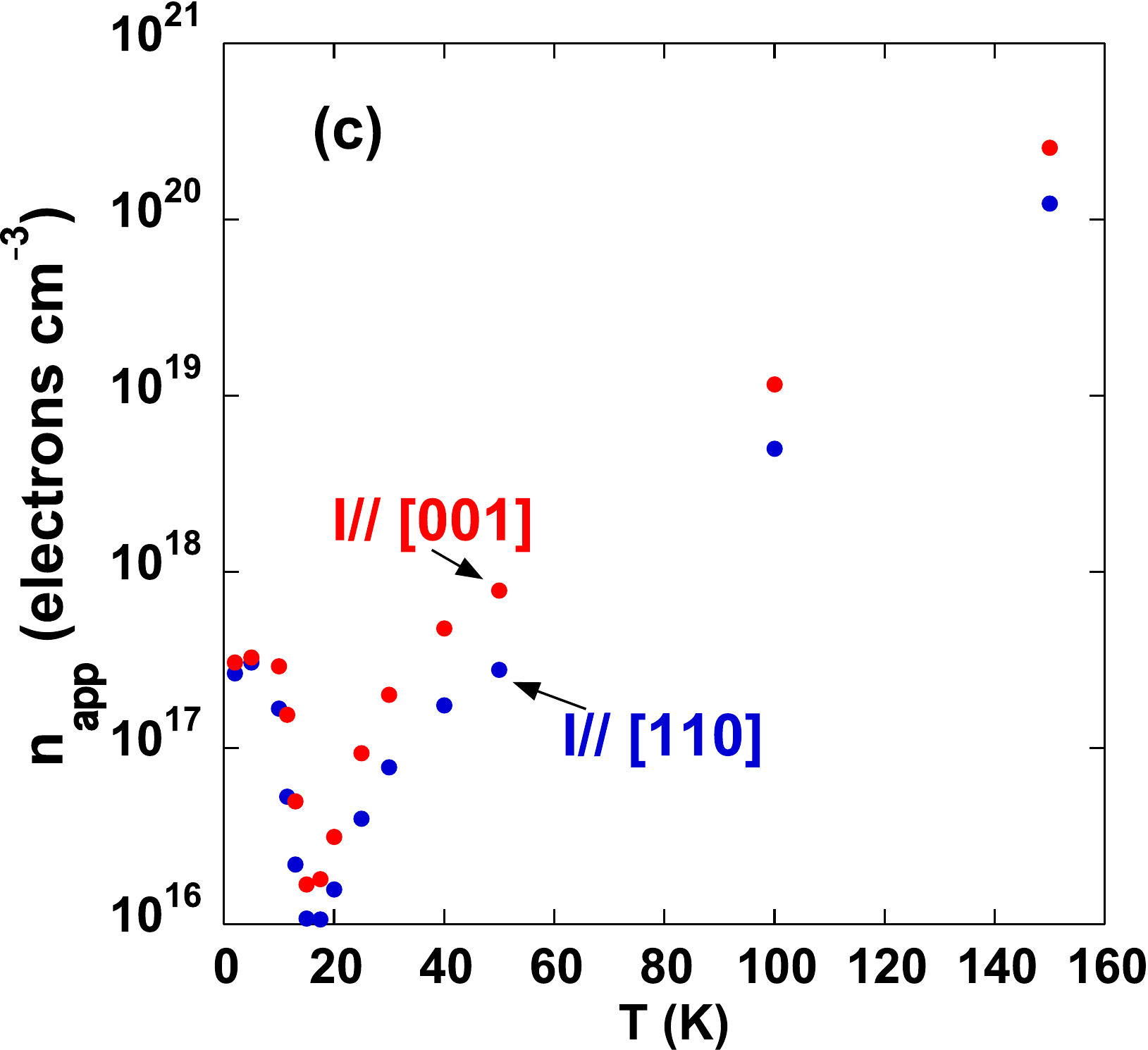}
\caption {(color online).(a) Hall resistivity versus magnetic field with the current along [110] and (b) along [001] for two crystals of CrSb$_2$. (c) Apparent carrier concentration versus temperature. Dashed lines are the linear fits to the data. Note minimum in apparent carrier concentration near 15\,K.}
\label{Hall}
\end{figure*}

Examples of the Hall resistivity ($\rho_{\mathrm{xy}}$) versus magnetic field for CrSb$_2$ crystals are shown in Fig.\ref{Hall}. All Hall resistivity data are linear in applied magnetic field at least up to fields of 4 Tesla, independent of whether the current is along [110] or [001]. We note that the Hall data for CrSb$_2$ are easier to analyze than for FeSb$_2$, since $\rho_{\mathrm{xy}}$ is linear in field for CrSb$_2$ at all temperatures where bipolar conduction is not present, unlike the complicated field dependence of $\rho_{\mathrm{xy}}$ exhibited by FeSb$_2$.\cite{Takahashi_Terasaki_2011} The most striking feature of the Hall data is the maximum that occurs in the magnitude of the Hall resistivity at T$\approx$15\,K, corresponding to an  apparent minimum in the carrier concentration. This is more evident when the apparent carrier concentration (determined from the one band expression n$_{\mathrm{app}}$$\propto$1/$\rho_{\mathrm{xy}}$) is plotted versus temperature (see Fig.\ref{Hall}c). The Hall and Seebeck signals are both negative below 250\,K. At high temperatures ($>$250\,K) large numbers of intrinsic electrons and holes are excited in CrSb$_2$, and the Hall signal is found to be positive while the Seebeck coefficient remains negative.

The minimum gap between the valence and conduction bands of CrSb$_2$ is estimated to be 140\,meV.\cite{Hu_2007,Harada_2004} This implies that intrinsic electron-hole pairs can be neglected for temperatures less than about 50\,K.  Below 50\,K, the apparent carrier concentration continues to decrease which suggests a freeze out of extrinsic carriers into a donor band.  The position of the donor level is estimated from the resistivity data to be in the range of 16\,meV below the conduction band edge.   The source of extrinsic electrons in CrSb$_2$ is unknown. A slight Sb deficiency ($\approx$0.00001) could account for the low temperature (5\,K) carrier concentration of about 3$\times$10$^{17}$\,electrons/cm$^3$, as could a small amount of electronically active impurities in the starting materials or from the alumina crucibles. A minimum in the apparent carrier concentration is unusual, but it has been observed in semiconductors such as Ge, \cite{Fritzsche_1955} as well as in FeSb$_2$ and FeAs$_2$.\cite{Sun_Dalton_2010} In Ge, the extrinsic carriers freeze out into a low mobility impurity band that is close to the conduction (or valence) band edge.  When the impurity level is within a few meV of the conduction or valence band edge, and the total concentration of dopants is in the range of 10$^{17}$\,carrier/cm$^3$, a low temperature minimum in the apparent carrier concentration can be observed.  It is important to stress that the carriers are not entirely localized in the donor band, though conduction in this band occurs with a low mobility and may be hopping-type conduction.  Indeed, the obtained temperature dependence of n$_{\mathrm{app}}$ originates in the dominance of the donor band conductivity at temperatures below the minimum in n$_{\mathrm{app}}$.



To determine the occupation of the dopant and conduction bands with increasing temperature, we use a simple model of two localized levels, where the occupation of conduction (band) levels is given by:

\begin{equation}
\mathrm{N_{con}} = 3\times10^{17}\frac{G \mathrm{exp}(-\Delta/k_{\mathrm{B}}T)}{1+G \mathrm{exp}(-\Delta/k_{\mathrm{B}}T)},
\label{eqn:twoband1}
\end{equation}

\noindent where $\Delta$ is the energy difference between the two levels, and $G$ is the degeneracy of the conduction level relative to the dopant level.  The energy of the dopant level is defined to be at E=0. The number of electrons remaining in the impurity band, N$_{\mathrm{imp}}$, is 3$\times$10$^{17}$ - N$_{\mathrm{con}}$. For temperatures less than about 50\,K, intrinsic conduction (the valence band) can be neglected, and the system is in the regime of k$_{\mathrm{B}}$T/$\Delta$ $<<$  1. We note that in this temperature regime the standard expression\cite{AshcroftMermin} for the number of electrons in the conduction band is  N$_{\mathrm{con}}$ $\approx$ N$_{\mathrm{c}}$exp(-$\Delta$/k$_{\mathrm{B}}$T), where N$_{\mathrm{c}}$ is the effective density of states of the conduction band, implying that in our simplified model $G$ is proportional to N$_{\mathrm{c}}$.

\begin{figure}[!ht]
\includegraphics [width=2.8in] {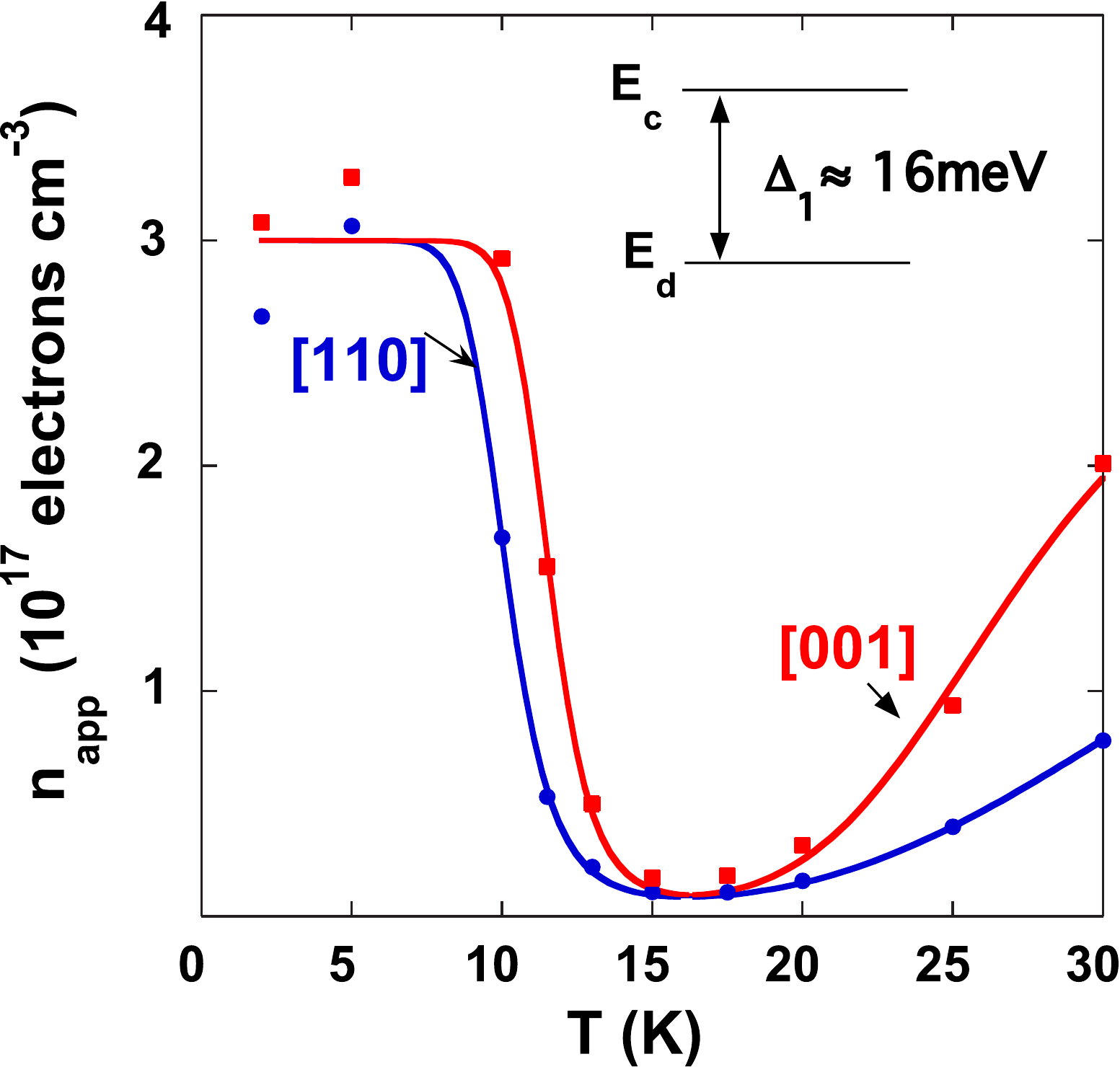}
\caption {(color online).Comparison between the measured apparent carrier concentration n$_{\mathrm{app}}$ with current along either [001] (solid squares) or [110] (solid circles) with the values calculated from two electron bands: a low mobility donor level, E$_{\mathrm{d}}$, in the gap a distance $\Delta$ below the conduction band edge at E$_{\mathrm{c}}$. A least squares fit using a simplified two-level model (Eqns.\,\ref{eqn:twoband1} and \ref{eqn:twoband2}) for transport along [110] yields $\Delta$ $\approx$ 12\,meV, $b$ $\approx$ 138, and $G$ = 29; for transport along [001], $\Delta$ $\approx$ 17.4\,meV, $b$ = 122 and $G$ = 1097.}
\label{napp}
\end{figure}

For the two band system, the apparent carrier concentration is given by
\begin{equation}
\mathrm{n_{app}} = \frac{(\mathrm{N_{imp}}+\mathrm{N_{con}}b)^2}{\mathrm{N_{imp}}+\mathrm{N_{con}}b^2}
\label{eqn:twoband2}
\end{equation}

\noindent where $b$ is ratio of the mobility of an electron in the conduction band to the value in the impurity band. This simple expression provides a good description of the Hall data, particularly for temperatures less than about 30\,K, as shown in Fig.\,\ref{napp}.  Although the exact fitting values given in the caption of Fig.\,\ref{napp} should not be taken too seriously because of the extreme simplicity of the model, they do indicate that the mobility of the conduction level is about 130 times higher than that of a donor level and the donor band lies $\approx$16\,meV below the conduction band.

Since CrSb$_2$ is antiferromagnetic, there is the possibility of an anomalous Hall effect contribution to the measured data.\cite{Kevane_1953,Paschen_2003} This would add a term to the inferred Hall coefficient which is proportional to the magnetic susceptibility. In the temperature range of interest (T$<$50\,K) the magnetic susceptibility of CrSb$_2$ is temperature independent for all orientations of the crystal with respect to the applied magnetic field, and all isothermal M versus H curves are linear up to the maximum field of 7 T. \cite{Hu_2007}  Any anomalous Hall effects would therefore result in at most a temperature independent offset in the Hall coefficient or inferred carrier concentration, and could not account for the observed minimum in Fig. \ref{napp}. The observation that the magnetic susceptibility varies by about 50\% between the two directions measured,\cite{Hu_2007} while the inferred carrier concentration at the lowest temperatures are essentially independent of orientation (Fig.\,\ref{napp}), suggests that anomalous Hall effects are negligible in this material at the temperatures where the Hall data are analyzed.

\begin{figure}[!ht]
\includegraphics [width=3in] {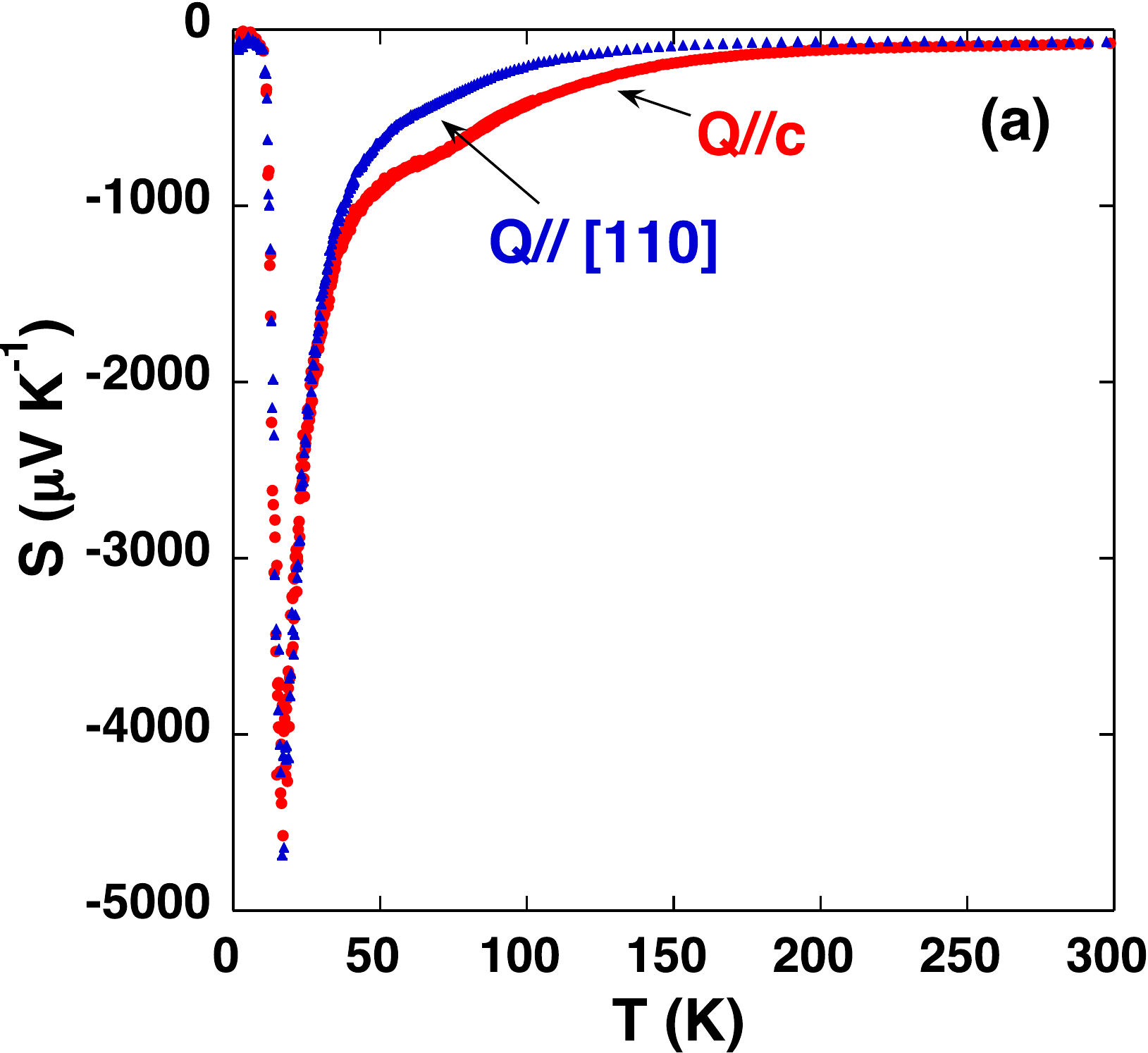}
\includegraphics [width=3in] {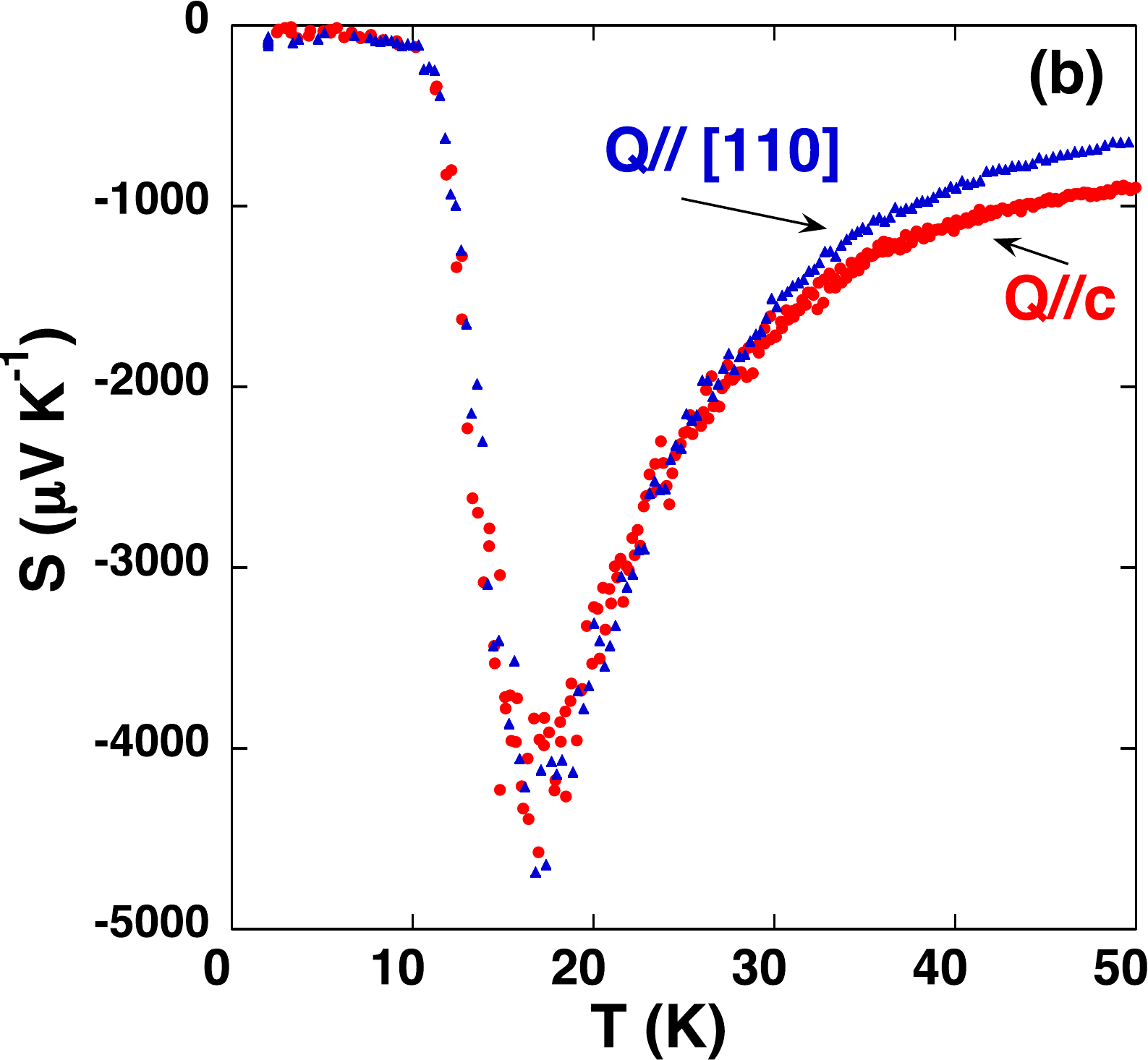}
\caption {(color online).Seebeck coefficient from CrSb$_2$ crystals with the temperature gradient along either [001] or [110]. Panel (b) shows the same data as in (a) but with an expanded temperature axis.}
\label{Seebeck}
\end{figure}

The most striking transport feature of CrSb$_2$ is the Seebeck coefficient, S, which is displayed in Fig.\,\ref{Seebeck}. The shape of S(T) is very similar to that previously reported for some FeSb$_2$ single crystals,\cite{Bentien_1994,Wang_2012} although there is some controversy as to how large S can be in FeSb$_2$. \cite{Takahashi_Sato_2011} In CrSb$_2$ single crystals, a minimum in S(T) of $\approx$ -4500\,$\mu$V/K occurs at 18\,K. From 18\,K to 10\,K there is a rapid decrease in the magnitude of S, while S changes much more slowly from 10-2\,K. The Seebeck data below 50\,K are very similar in shape to the Hall data from the same temperature range (see Fig\,\ref{napp}). For a two band system the Seebeck coefficient is given by the Seebeck coefficient of each band weighted by the electrical conductivity of each band.\cite{Goldsmid_1986,Sales_2010}. The rapid drop in S below 18 K can be accounted for by the thermal depopulation of electrons from the non-degenerate conduction band into the low mobility and degenerate donor band upon cooling.  Below 10\,K, the Seebeck coefficient is approximately linear in temperature and is dominated by donor band conduction.

The large magnitude of S at 18\,K is difficult to reconcile with a purely electronic mechanism, particularly since the Cr moments order at T$_{\mathrm{N}}$ $\approx$ 273\,K and hence there is no additional magnetic entropy to be removed at lower temperatures. The heat capacity and lattice thermal conductivity data, which are discussed below, are consistent with a phonon drag mechanism as the origin of the large values for S near 20\,K. We note that the CrSb$_2$ Seebeck data were reproducible using crystals from different growths and by measuring the Seebeck data using different heating and cooling parameters. The very similar shape between the Seebeck data from FeSb$_2$\cite{Bentien_1994,Wang_2012} and the Seebeck data shown in Fig.\,\ref{Seebeck}, suggests a similar origin for the effect.  

\begin{figure}[!ht]
\includegraphics [width=3in] {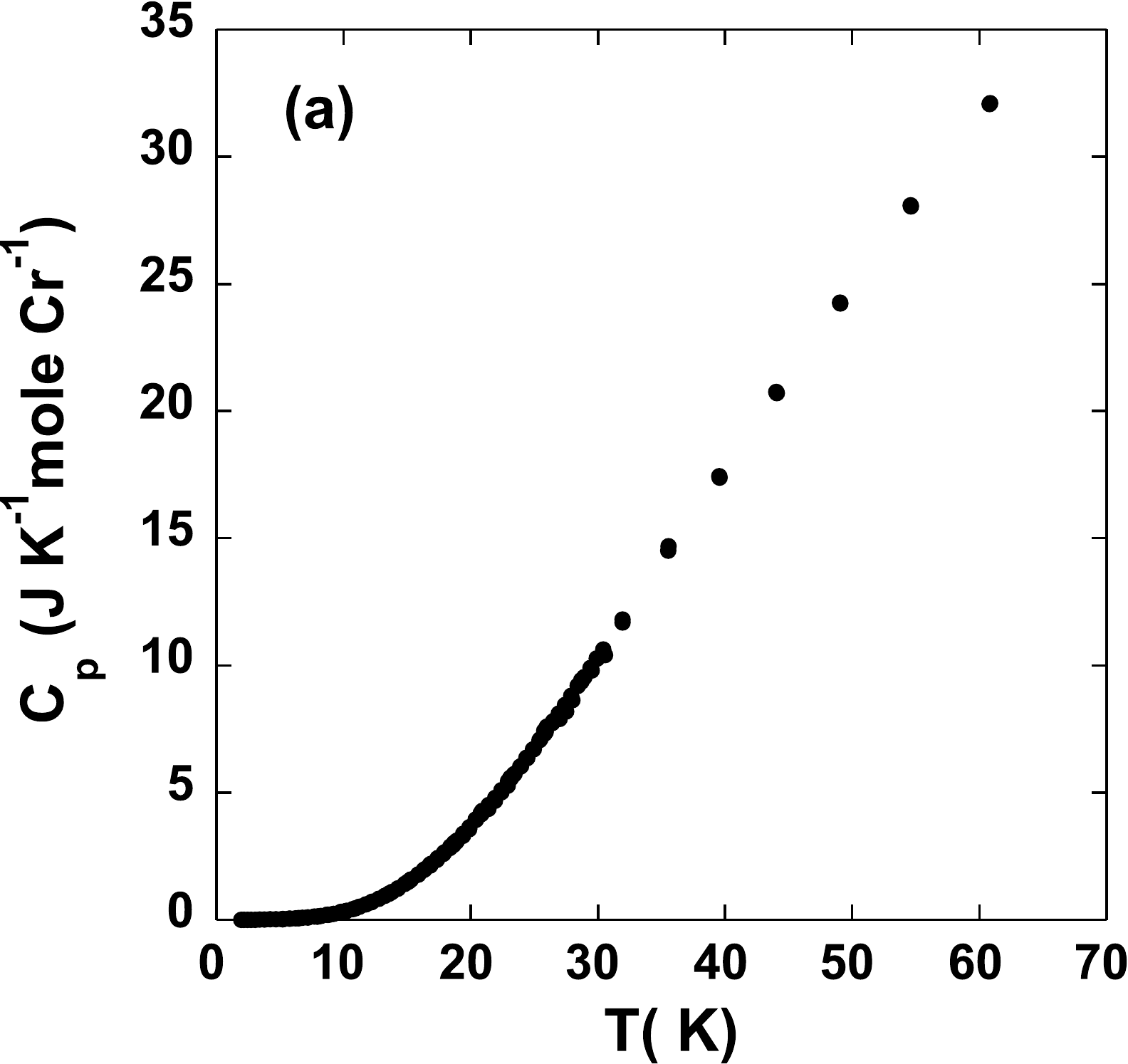}
\includegraphics [width=3in] {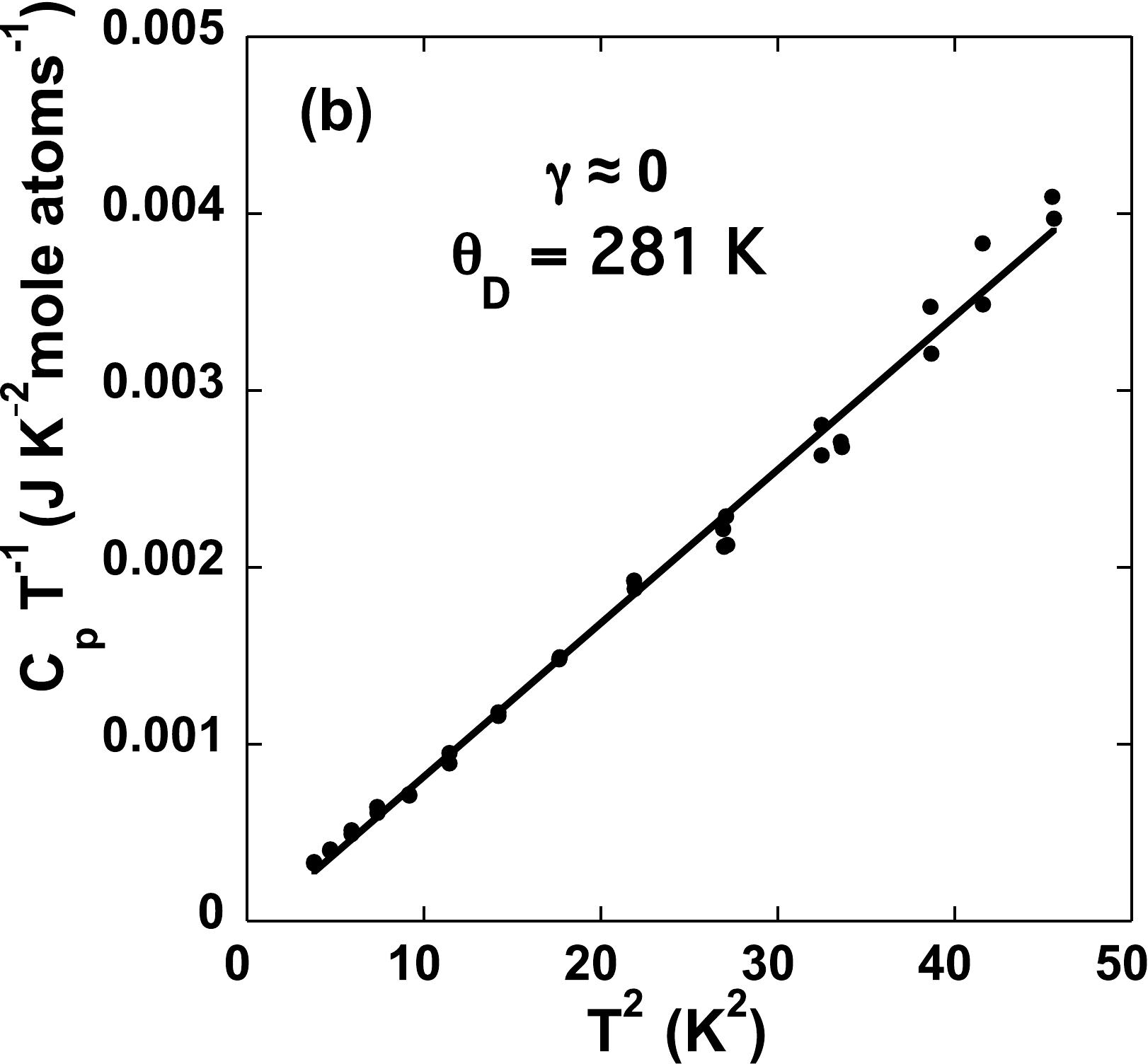}
\caption {(color online). (a) Heat capacity vs temperature for CrSb$_2$ crystal, and in (b) the same data are plotted as C/T vs T$^2$ below 7\,K.}
\label{Cp}
\end{figure}

The low temperature heat capacity data for CrSb$_2$ are shown in Fig.\,\ref{Cp}. As noted above and by previous work,\cite{Holseth_1970,Alles_1978,Stone_2012} there are no phase transitions at low temperatures except for that at T$_{\mathrm{N}}$ $\approx$ 273\,K. Plots of C/T vs T are linear for temperatures less than 7\,K and as expected for a semiconductor, the electronic specific heat coefficient, $\gamma \approx$ 0, within experimental error.  The Debye temperature is 281\,K, which implies an average sound velocity of v$_{\mathrm{s}}$ = 2700\,m/s.  Although CrSb$_2$ is magnetically ordered below T$_{\mathrm{N}}$= 273\,K, at low temperatures there is a substantial gap in the spin wave spectrum of about 25\,meV.\cite{Stone_2012} Thus the heat capacity below 7\,K is only due to phonons.

\begin{figure}[!ht]
\includegraphics [width=3in] {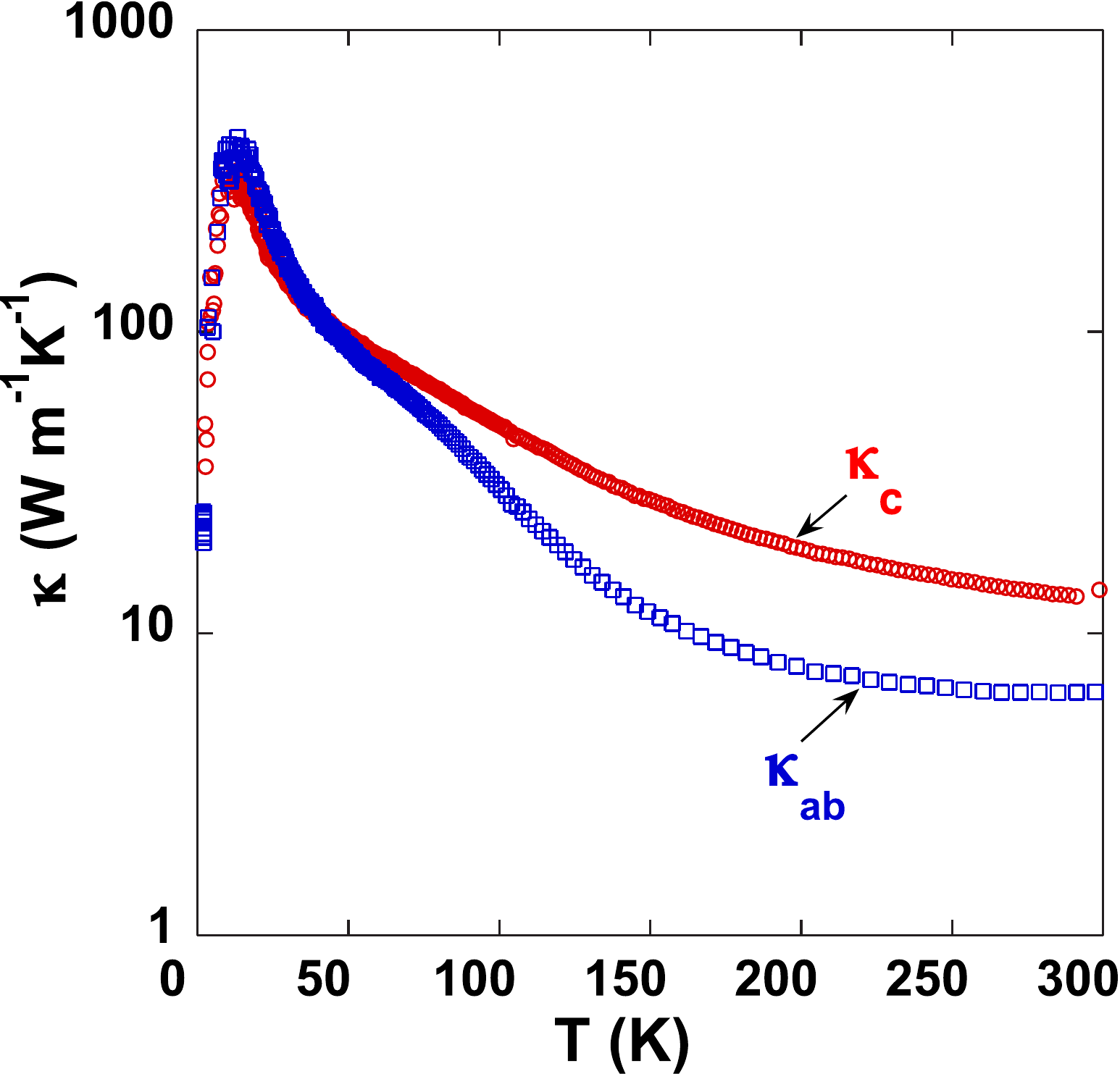}
\caption {(color online) Thermal conductivity versus temperature for two CrSb$_2$ crystals. The thermal conductivity was measured along the $c$ direction and perpendicular to $c$ (along [110]).}
\label{kappa}
\end{figure}

The thermal conductivity data for two CrSb$_2$ single crystals are shown in Fig.\,\ref{kappa} for heat flow along the $c$ direction, $\kappa_{\mathrm{c}}$, and along [110], $\kappa_{\mathrm{ab}}$. For temperatures below 50\,K, the thermal conductivity data are very similar in both directions with a large maximum of 300-400\,W/m-K near 12\,K. This is approximately a factor of 10 larger than that observed in polycrystalline CrSb$_2$.\cite{Li_2008} This implies that the average mean free path for phonons at low temperature is about 10 times smaller in the polycrystalline samples. For the polycrystalline samples \cite{Li_2008} the peak in the Seebeck data is also a factor of 10 smaller. All of these observations are consistent with a phonon drag mechanism as the origin of the large peak in the Seebeck data near 18 K, as is discussed in more detail below. The electronic contribution to the thermal conductivity is estimated to be negligible at all temperatures using the Wiedemann-Franz law and the resistivity data shown in Fig.\,\ref{resist} The effect of subtracting the electronic contribution to the data shown in Fig.\,\ref{kappa} is less than the size of the data symbols used in the figure.

Above 50\,K, $\kappa_{\mathrm{c}}$ is substantially larger than $\kappa_{\mathrm{ab}}$. The magnetic excitations in CrSb$_2$ are quasi-1D with high spin-wave velocities along the $c$ direction.   It is possible that the excess heat carried along the $c$ direction is due to spin-waves since in this direction the spin wave velocities are about three times larger than the average sound velocity for CrSb$_2$. The gap of 25\,meV in the spin-wave spectrum at the zone center is also consistent with an onset of spin-wave heat conduction near 50\,K. However, since $\kappa_{\mathrm{c}}$ is not an order of magnitude larger than $\kappa_{\mathrm{ab}}$, it is also possible that the magnetic excitations scatter phonons more strongly for $\kappa_{\mathrm{ab}}$ thereby suppressing $\kappa_{\mathrm{ab}}$ relative to $\kappa_{\mathrm{c}}$. There is no simple way to distinguish between these two possibilities for CrSb$_2$. Substantial heat conduction by magnetic excitations has been clearly demonstrated in the insulating spin-ladder compound Ca$_9$La$_5$Cu$_{24}$O$_{41}$. Along the ladder direction the heat conducted by spin-waves at room temperature is comparable to the heat conducted by metallic platinum.\cite{Hess_2001} In the spin-ladder compound, however, it was found that a small amount of electronic conduction tended to greatly reduce the spin-wave heat conduction due to electron-magnon scattering. It is likely that the presence of conduction electrons in CrSb$_2$ has a similar effect.

A quantitative estimate of a phonon drag contribution to the Seebeck coefficient is difficult.  An approximate estimate\cite{Takahashi_Terasaki_2011,Weber_1991,Herring_1954} for the phonon drag contribution to S is given by S$_{\mathrm{ph}}$=$\beta$v$_{\mathrm{s}}$l$_{\mathrm{p}}$/$\mu$T, where $\beta$ is a parameter that characterizes the relative strength of the electron-phonon interaction with 0$<\beta<$1, and $\mu$ is the mobility of electrons in the conduction band.  Using the measured low temperature values for the heat capacity and thermal conductivity, the estimated mean free path for phonons, l$_{\mathrm{p}}$, is about 7\,$\mu$m at 18\,K. Using the simple two band analysis of the Hall data, presented above, yields a value of $\mu$ $\sim$ 500-2500\,cm$^2$/V-s at 18\,K depending on whether the current is along [110] or along the $c$ axis. This yields S$_{\mathrm{ph}}$=$\beta$(20-100)\,mV/K at 18\,K, which indicates that phonon drag could account for a large percentage of the negative peak in the Seebeck data shown in Fig.\,\ref{Seebeck}.  A phonon drag explanation is also consistent with the S(T) observed in polycrystalline CrSb$_2$, where a maximum $|$S$|$\,$\sim$450\,$\mu$V/K is observed near 60\,K and corresponds to a smaller maximum in $\kappa$ of approximately 30\,W/m/K at similar T.  
The very similar shape between the Seebeck data from FeSb$_2$\cite{Bentien_1994,Wang_2012} and the Seebeck data shown in Fig.\,\ref{Seebeck}, suggests a similar origin for the effect.

We note that the high values of the Seebeck coefficient cannot be explained in terms of the behavior of a standard doped semiconductor model with temperature independent doping in the absence of phonon drag or some other enhancement.\cite{Kuroki_2007} This is because, although the thermopower may increase as one decreases the carrier concentration, this increase will be limited by bi-polar conduction at a finite temperature. While the details will in general depend on the band structure, Goldsmid and Sharp\cite{Sharp_1999} derived an approximate expression for the maximum thermopower of a band material, S$_{\mathrm{max}}$ $\approx$ E$_{\mathrm{g}}$/2eT$_{\mathrm{max}}$. Here E$_{\mathrm{g}}$ is the band gap and T$_{\mathrm{max}}$ is the temperature where the maximum thermopower is observed. For 18\,K and the literature band gap of 140\,meV, this formula gives S$_{\mathrm{max}}$ $\approx$ -3880$\mu$V/K, which is not too far from the observed value of -4500$\mu$V/K. However, the carrier concentration needed to achieve this maximum value at this T is several orders of magnitude smaller than found experimentally. If the more relevant low temperature gap of 16\,meV is used, however, S$_{\mathrm{max}}$ $\sim$ 450\,$\mu$V/K, which is about 10 times smaller than the experimental value for any carrier concentration. While it is apparent that with a substantial gap one can get very large thermopowers at low T, unless the bands are very heavy this only happens at exceedingly low carrier concentrations.

We performed first principles calculations of the electronic structure using the general potential linearized augmented planewave method as implemented in the WIEN2k code.\cite{wien} Calculations were done for the experimentally determined ground state magnetic structure, with the standard Perdew-Burke-Ernzerhof (PBE) exchange correlation functional.

We find a very small band gap of 5.4\,meV in scalar relativistic calculations with the PBE functional and a band gap that depends on the moment direction if spin-orbit is included. We obtain a somewhat larger gap of 70\,meV in scalar relativistic calculations using the local-spin-density approximation. For the experimental direction we obtain a zero gap with spin-orbit and the PBE functional. This underestimate and the differences between different functionals probably reflect the known limitations of standard density functional calculations. In order to connect with experiment we calculated the Seebeck coefficient as a function of doping level and temperature within the constant scattering time approximation, as described elsewhere.\cite{Boltztrap,Madsen_2003} We did this for the PBE electronic structure obtained from scalar relativistic calculations with application of a scissors operator to adjust the gap to 16\,meV and 140\,meV. We find that the calculated electronic structure cannot account for the large Seebeck coefficient peak of -4500$\mu$V/K at 18\,K.

Our calculations, however, did find an interesting result that could be checked by future experiments. As shown in Ref.\citenum{Stone_2012}, even though the band gap is small, the shape of the density of states for the valence and conduction bands is quite different. The valence band edge is heavier than the conduction band edge. Furthermore the density of states at the top of the valence bands is from hybridized Cr $d$ / Sb $p$ states, while the conduction band has more pure Cr $d$ character. The predicted behavior for p-type and n-type CrSb$_2$ are therefore different. For n-type, we find a low temperature conductivity anisotropy within the constant scattering time approximation of $\sim$6.5, with the high conductivity along the $c$-axis, consistent with experiment. For p-type, the anisotropy is predicted to be $\sim$4 with lowest conductivity along the orthorhombic $a$ axis, opposite to the n-type case.

\section{Summary and Conclusions}
The low temperature electrical and thermal transport properties of single crystals of the narrow-gap magnetic semiconductor (T$_{\mathrm{N}}\approx$273\,K) CrSb$_2$ are reported. A large negative peak in the Seebeck coefficient of -4500$\mu$V/K is observed at 18\,K with heat flow either along the $c$ axis or perpendicular to the $c$ axis. The Hall resistivity is linear in applied magnetic fields up to 4\,T for all crystallographic orientations studied and all temperatures where bipolar conduction is not influential. The Hall and Seebeck coefficients are both negative for temperatures below 250 K. The sharp maximum in the magnitude of the Hall resistivity for temperatures near 18\,K clearly suggests two electron bands influence electrical transport at low T. The resistivity and Hall data are consistent with an occupied low-mobility donor band at 2\,K with the conduction band edge about 16\,meV higher in energy. The thermal conductivity and heat capacity data are consistent with a significant phonon drag contribution to the large negative value of the Seebeck coefficient at 18\,K. We find that the calculated electronic structure cannot account for the large Seebeck peak. While CrSb$_2$ is magnetic and the states near the conduction and valence band edges have strong 3$d$ character, most of the low temperature properties, including the Seebeck data, can be understood within the framework of a relatively normal narrow gap semiconductor. The temperature dependence of the Seebeck coefficient for CrSb$_2$ is remarkably similar in shape to that reported for FeSb$_2$.\cite{Bentien_1994,Wang_2012}

\section{Acknowledgements}
Research supported by the U.S. Department of Energy, Basic Energy Sciences, Materials Sciences and Engineering Division (B.C.S., A.F.M., M.A.M., D.J.S., and D.M.), and by the Department of Energy, Office of Scientific User Facilities (M.B.S.).


\begin{thebibliography}{41}%
\makeatletter
\providecommand \@ifxundefined [1]{%
 \@ifx{#1\undefined}
}%
\providecommand \@ifnum [1]{%
 \ifnum #1\expandafter \@firstoftwo
 \else \expandafter \@secondoftwo
 \fi
}%
\providecommand \@ifx [1]{%
 \ifx #1\expandafter \@firstoftwo
 \else \expandafter \@secondoftwo
 \fi
}%
\providecommand \natexlab [1]{#1}%
\providecommand \enquote  [1]{``#1''}%
\providecommand \bibnamefont  [1]{#1}%
\providecommand \bibfnamefont [1]{#1}%
\providecommand \citenamefont [1]{#1}%
\providecommand \href@noop [0]{\@secondoftwo}%
\providecommand \href [0]{\begingroup \@sanitize@url \@href}%
\providecommand \@href[1]{\@@startlink{#1}\@@href}%
\providecommand \@@href[1]{\endgroup#1\@@endlink}%
\providecommand \@sanitize@url [0]{\catcode `\\12\catcode `\$12\catcode
  `\&12\catcode `\#12\catcode `\^12\catcode `\_12\catcode `\%12\relax}%
\providecommand \@@startlink[1]{}%
\providecommand \@@endlink[0]{}%
\providecommand \url  [0]{\begingroup\@sanitize@url \@url }%
\providecommand \@url [1]{\endgroup\@href {#1}{\urlprefix }}%
\providecommand \urlprefix  [0]{URL }%
\providecommand \Eprint [0]{\href }%
\providecommand \doibase [0]{http://dx.doi.org/}%
\providecommand \selectlanguage [0]{\@gobble}%
\providecommand \bibinfo  [0]{\@secondoftwo}%
\providecommand \bibfield  [0]{\@secondoftwo}%
\providecommand \translation [1]{[#1]}%
\providecommand \BibitemOpen [0]{}%
\providecommand \bibitemStop [0]{}%
\providecommand \bibitemNoStop [0]{.\EOS\space}%
\providecommand \EOS [0]{\spacefactor3000\relax}%
\providecommand \BibitemShut  [1]{\csname bibitem#1\endcsname}%
\let\auto@bib@innerbib\@empty
\bibitem [{\citenamefont {Jaccarino}\ \emph {et~al.}(1967)\citenamefont
  {Jaccarino}, \citenamefont {Wertheim}, \citenamefont {Wernick}, \citenamefont
  {Walker},\ and\ \citenamefont {Arajs}}]{Jaccarino_1976}%
  \BibitemOpen
  \bibfield  {author} {\bibinfo {author} {\bibfnamefont {V.}~\bibnamefont
  {Jaccarino}}, \bibinfo {author} {\bibfnamefont {G.~K.}\ \bibnamefont
  {Wertheim}}, \bibinfo {author} {\bibfnamefont {J.~H.}\ \bibnamefont
  {Wernick}}, \bibinfo {author} {\bibfnamefont {L.~R.}\ \bibnamefont {Walker}},
  \ and\ \bibinfo {author} {\bibfnamefont {S.}~\bibnamefont {Arajs}},\ }\href
  {\doibase 10.1103/PhysRev.160.476} {\bibfield  {journal} {\bibinfo  {journal}
  {Phys. Rev.}\ }\textbf {\bibinfo {volume} {160}},\ \bibinfo {pages} {476}
  (\bibinfo {year} {1967})}\BibitemShut {NoStop}%
\bibitem [{\citenamefont {Mandrus}\ \emph {et~al.}(1995)\citenamefont
  {Mandrus}, \citenamefont {Sarrao}, \citenamefont {Migliori}, \citenamefont
  {Thompson},\ and\ \citenamefont {Fisk}}]{Mandrus_1995_FeSi}%
  \BibitemOpen
  \bibfield  {author} {\bibinfo {author} {\bibfnamefont {D.}~\bibnamefont
  {Mandrus}}, \bibinfo {author} {\bibfnamefont {J.~L.}\ \bibnamefont {Sarrao}},
  \bibinfo {author} {\bibfnamefont {A.}~\bibnamefont {Migliori}}, \bibinfo
  {author} {\bibfnamefont {J.~D.}\ \bibnamefont {Thompson}}, \ and\ \bibinfo
  {author} {\bibfnamefont {Z.}~\bibnamefont {Fisk}},\ }\href {\doibase
  10.1103/PhysRevB.51.4763} {\bibfield  {journal} {\bibinfo  {journal} {Phys.
  Rev. B}\ }\textbf {\bibinfo {volume} {51}},\ \bibinfo {pages} {4763}
  (\bibinfo {year} {1995})}\BibitemShut {NoStop}%
\bibitem [{\citenamefont {Sales}\ \emph {et~al.}(1994)\citenamefont {Sales},
  \citenamefont {Jones}, \citenamefont {Chakoumakos}, \citenamefont
  {Fernandez-Baca}, \citenamefont {Harmon}, \citenamefont {Sharp},\ and\
  \citenamefont {Volckmann}}]{Sales_1994_FeIrSi}%
  \BibitemOpen
  \bibfield  {author} {\bibinfo {author} {\bibfnamefont {B.~C.}\ \bibnamefont
  {Sales}}, \bibinfo {author} {\bibfnamefont {E.~C.}\ \bibnamefont {Jones}},
  \bibinfo {author} {\bibfnamefont {B.~C.}\ \bibnamefont {Chakoumakos}},
  \bibinfo {author} {\bibfnamefont {J.~A.}\ \bibnamefont {Fernandez-Baca}},
  \bibinfo {author} {\bibfnamefont {H.~E.}\ \bibnamefont {Harmon}}, \bibinfo
  {author} {\bibfnamefont {J.~W.}\ \bibnamefont {Sharp}}, \ and\ \bibinfo
  {author} {\bibfnamefont {E.~H.}\ \bibnamefont {Volckmann}},\ }\href {\doibase
  10.1103/PhysRevB.50.8207} {\bibfield  {journal} {\bibinfo  {journal} {Phys.
  Rev. B}\ }\textbf {\bibinfo {volume} {50}},\ \bibinfo {pages} {8207}
  (\bibinfo {year} {1994})}\BibitemShut {NoStop}%
\bibitem [{\citenamefont {Wolfe}\ \emph {et~al.}()\citenamefont {Wolfe},
  \citenamefont {Wernick},\ and\ \citenamefont {Haszko}}]{Wolfe_1965}%
  \BibitemOpen
  \bibfield  {author} {\bibinfo {author} {\bibfnamefont {R.}~\bibnamefont
  {Wolfe}}, \bibinfo {author} {\bibfnamefont {J.~H.}\ \bibnamefont {Wernick}},
  \ and\ \bibinfo {author} {\bibfnamefont {S.~E.}\ \bibnamefont {Haszko}},\
  }\href@noop {} {\bibfield  {journal} {\bibinfo  {journal} {Phys. Lett.}\
  }\textbf {\bibinfo {volume} {19}},\ \bibinfo {pages} {449}}\BibitemShut
  {NoStop}%
\bibitem [{\citenamefont {DeGiorgi}\ \emph {et~al.}(1994)\citenamefont
  {DeGiorgi}, \citenamefont {Hunt}, \citenamefont {Ott}, \citenamefont
  {Dressel}, \citenamefont {Feenstra}, \citenamefont
  {Gr$\ddot{\mathrm{u}}$ner}, \citenamefont {Fisk},\ and\ \citenamefont
  {Canfield}}]{DeGiorgi_1994}%
  \BibitemOpen
  \bibfield  {author} {\bibinfo {author} {\bibfnamefont {L.}~\bibnamefont
  {DeGiorgi}}, \bibinfo {author} {\bibfnamefont {M.~B.}\ \bibnamefont {Hunt}},
  \bibinfo {author} {\bibfnamefont {H.~R.}\ \bibnamefont {Ott}}, \bibinfo
  {author} {\bibfnamefont {M.}~\bibnamefont {Dressel}}, \bibinfo {author}
  {\bibfnamefont {B.~J.}\ \bibnamefont {Feenstra}}, \bibinfo {author}
  {\bibfnamefont {G.}~\bibnamefont {Gr$\ddot{\mathrm{u}}$ner}}, \bibinfo
  {author} {\bibfnamefont {Z.}~\bibnamefont {Fisk}}, \ and\ \bibinfo {author}
  {\bibfnamefont {P.}~\bibnamefont {Canfield}},\ }\href@noop {} {\bibfield
  {journal} {\bibinfo  {journal} {Europhys. Lett.}\ }\textbf {\bibinfo {volume}
  {28}},\ \bibinfo {pages} {341} (\bibinfo {year} {1994})}\BibitemShut
  {NoStop}%
\bibitem [{\citenamefont {Onose}\ \emph {et~al.}(2005)\citenamefont {Onose},
  \citenamefont {Takeshita}, \citenamefont {Terakura}, \citenamefont {Takagi},\
  and\ \citenamefont {Tokura}}]{Onose_Tokura_2005}%
  \BibitemOpen
  \bibfield  {author} {\bibinfo {author} {\bibfnamefont {Y.}~\bibnamefont
  {Onose}}, \bibinfo {author} {\bibfnamefont {N.}~\bibnamefont {Takeshita}},
  \bibinfo {author} {\bibfnamefont {C.}~\bibnamefont {Terakura}}, \bibinfo
  {author} {\bibfnamefont {H.}~\bibnamefont {Takagi}}, \ and\ \bibinfo {author}
  {\bibfnamefont {Y.}~\bibnamefont {Tokura}},\ }\href {\doibase
  10.1103/PhysRevB.72.224431} {\bibfield  {journal} {\bibinfo  {journal} {Phys.
  Rev. B}\ }\textbf {\bibinfo {volume} {72}},\ \bibinfo {pages} {224431}
  (\bibinfo {year} {2005})}\BibitemShut {NoStop}%
\bibitem [{\citenamefont {Bentien}\ \emph {et~al.}(2007)\citenamefont
  {Bentien}, \citenamefont {Johnsen}, \citenamefont {Madsen}, \citenamefont
  {Iversen},\ and\ \citenamefont {Steglich}}]{Bentien_1994}%
  \BibitemOpen
  \bibfield  {author} {\bibinfo {author} {\bibfnamefont {A.}~\bibnamefont
  {Bentien}}, \bibinfo {author} {\bibfnamefont {S.}~\bibnamefont {Johnsen}},
  \bibinfo {author} {\bibfnamefont {G.~K.~H.}\ \bibnamefont {Madsen}}, \bibinfo
  {author} {\bibfnamefont {B.~B.}\ \bibnamefont {Iversen}}, \ and\ \bibinfo
  {author} {\bibfnamefont {F.}~\bibnamefont {Steglich}},\ }\href@noop {}
  {\bibfield  {journal} {\bibinfo  {journal} {E P L}\ }\textbf {\bibinfo
  {volume} {80}},\ \bibinfo {pages} {17008} (\bibinfo {year}
  {2007})}\BibitemShut {NoStop}%
\bibitem [{\citenamefont {Petrovic}\ \emph {et~al.}(2003)\citenamefont
  {Petrovic}, \citenamefont {Kim}, \citenamefont {Bud'ko}, \citenamefont
  {Goldman}, \citenamefont {Canfield}, \citenamefont {Choe},\ and\
  \citenamefont {Miller}}]{Petrovic_2003}%
  \BibitemOpen
  \bibfield  {author} {\bibinfo {author} {\bibfnamefont {C.}~\bibnamefont
  {Petrovic}}, \bibinfo {author} {\bibfnamefont {J.~W.}\ \bibnamefont {Kim}},
  \bibinfo {author} {\bibfnamefont {S.~L.}\ \bibnamefont {Bud'ko}}, \bibinfo
  {author} {\bibfnamefont {A.~I.}\ \bibnamefont {Goldman}}, \bibinfo {author}
  {\bibfnamefont {P.~C.}\ \bibnamefont {Canfield}}, \bibinfo {author}
  {\bibfnamefont {W.}~\bibnamefont {Choe}}, \ and\ \bibinfo {author}
  {\bibfnamefont {G.~J.}\ \bibnamefont {Miller}},\ }\href {\doibase
  10.1103/PhysRevB.67.155205} {\bibfield  {journal} {\bibinfo  {journal} {Phys.
  Rev. B}\ }\textbf {\bibinfo {volume} {67}},\ \bibinfo {pages} {155205}
  (\bibinfo {year} {2003})}\BibitemShut {NoStop}%
\bibitem [{\citenamefont {Takahashi}\ \emph
  {et~al.}(2011{\natexlab{a}})\citenamefont {Takahashi}, \citenamefont
  {Okazaki}, \citenamefont {Yasui},\ and\ \citenamefont
  {Terasaki}}]{Takahashi_Terasaki_2011}%
  \BibitemOpen
  \bibfield  {author} {\bibinfo {author} {\bibfnamefont {H.}~\bibnamefont
  {Takahashi}}, \bibinfo {author} {\bibfnamefont {R.}~\bibnamefont {Okazaki}},
  \bibinfo {author} {\bibfnamefont {Y.}~\bibnamefont {Yasui}}, \ and\ \bibinfo
  {author} {\bibfnamefont {I.}~\bibnamefont {Terasaki}},\ }\href {\doibase
  10.1103/PhysRevB.84.205215} {\bibfield  {journal} {\bibinfo  {journal} {Phys.
  Rev. B}\ }\textbf {\bibinfo {volume} {84}},\ \bibinfo {pages} {205215}
  (\bibinfo {year} {2011}{\natexlab{a}})}\BibitemShut {NoStop}%
\bibitem [{\citenamefont {Sun}\ \emph {et~al.}(2011)\citenamefont {Sun},
  \citenamefont {S{\o}ndergaard}, \citenamefont {Iversen},\ and\ \citenamefont
  {Steglich}}]{Sun_2011}%
  \BibitemOpen
  \bibfield  {author} {\bibinfo {author} {\bibfnamefont {P.}~\bibnamefont
  {Sun}}, \bibinfo {author} {\bibfnamefont {M.}~\bibnamefont {S{\o}ndergaard}},
  \bibinfo {author} {\bibfnamefont {B.~B.}\ \bibnamefont {Iversen}}, \ and\
  \bibinfo {author} {\bibfnamefont {F.}~\bibnamefont {Steglich}},\ }\href@noop
  {} {\bibfield  {journal} {\bibinfo  {journal} {Ann. Phys.}\ }\textbf
  {\bibinfo {volume} {523}},\ \bibinfo {pages} {612} (\bibinfo {year}
  {2011})}\BibitemShut {NoStop}%
\bibitem [{\citenamefont {Holseth}\ \emph {et~al.}(1970)\citenamefont
  {Holseth}, \citenamefont {Kjekshus},\ and\ \citenamefont
  {Andresen}}]{Holseth_1970}%
  \BibitemOpen
  \bibfield  {author} {\bibinfo {author} {\bibfnamefont {H.}~\bibnamefont
  {Holseth}}, \bibinfo {author} {\bibfnamefont {A.}~\bibnamefont {Kjekshus}}, \
  and\ \bibinfo {author} {\bibfnamefont {A.~F.}\ \bibnamefont {Andresen}},\
  }\href@noop {} {\bibfield  {journal} {\bibinfo  {journal} {Acta Chem.
  Scand.}\ }\textbf {\bibinfo {volume} {24}},\ \bibinfo {pages} {3309}
  (\bibinfo {year} {1970})}\BibitemShut {NoStop}%
\bibitem [{\citenamefont {Alles}\ \emph {et~al.}(1978)\citenamefont {Alles},
  \citenamefont {Falk}, \citenamefont {WestrumJr},\ and\ \citenamefont
  {Gr{\o}nvold}}]{Alles_1978}%
  \BibitemOpen
  \bibfield  {author} {\bibinfo {author} {\bibfnamefont {A.}~\bibnamefont
  {Alles}}, \bibinfo {author} {\bibfnamefont {B.}~\bibnamefont {Falk}},
  \bibinfo {author} {\bibfnamefont {E.~F.}\ \bibnamefont {WestrumJr}}, \ and\
  \bibinfo {author} {\bibfnamefont {F.}~\bibnamefont {Gr{\o}nvold}},\
  }\href@noop {} {\bibfield  {journal} {\bibinfo  {journal} {J. Chem. Therm.}\
  }\textbf {\bibinfo {volume} {10}},\ \bibinfo {pages} {103} (\bibinfo {year}
  {1978})}\BibitemShut {NoStop}%
\bibitem [{\citenamefont {Hu}\ \emph {et~al.}(2007)\citenamefont {Hu},
  \citenamefont {Mitrovi\ifmmode~\acute{c}\else \'{c}\fi{}},\ and\
  \citenamefont {Petrovic}}]{Hu_2007}%
  \BibitemOpen
  \bibfield  {author} {\bibinfo {author} {\bibfnamefont {R.}~\bibnamefont
  {Hu}}, \bibinfo {author} {\bibfnamefont {V.~F.}\ \bibnamefont
  {Mitrovi\ifmmode~\acute{c}\else \'{c}\fi{}}}, \ and\ \bibinfo {author}
  {\bibfnamefont {C.}~\bibnamefont {Petrovic}},\ }\href {\doibase
  10.1103/PhysRevB.76.115105} {\bibfield  {journal} {\bibinfo  {journal} {Phys.
  Rev. B}\ }\textbf {\bibinfo {volume} {76}},\ \bibinfo {pages} {115105}
  (\bibinfo {year} {2007})}\BibitemShut {NoStop}%
\bibitem [{\citenamefont {Delaire}\ \emph {et~al.}(2011)\citenamefont
  {Delaire}, \citenamefont {Marty}, \citenamefont {Stone}, \citenamefont
  {Kent}, \citenamefont {Lucas}, \citenamefont {Abernathy}, \citenamefont
  {Mandrus},\ and\ \citenamefont {Sales}}]{Delaire_2011}%
  \BibitemOpen
  \bibfield  {author} {\bibinfo {author} {\bibfnamefont {O.}~\bibnamefont
  {Delaire}}, \bibinfo {author} {\bibfnamefont {K.}~\bibnamefont {Marty}},
  \bibinfo {author} {\bibfnamefont {M.~B.}\ \bibnamefont {Stone}}, \bibinfo
  {author} {\bibfnamefont {P.~R.~C.}\ \bibnamefont {Kent}}, \bibinfo {author}
  {\bibfnamefont {M.~S.}\ \bibnamefont {Lucas}}, \bibinfo {author}
  {\bibfnamefont {D.~L.}\ \bibnamefont {Abernathy}}, \bibinfo {author}
  {\bibfnamefont {D.}~\bibnamefont {Mandrus}}, \ and\ \bibinfo {author}
  {\bibfnamefont {B.~C.}\ \bibnamefont {Sales}},\ }\href@noop {} {\bibfield
  {journal} {\bibinfo  {journal} {PNAS}\ }\textbf {\bibinfo {volume} {108}},\
  \bibinfo {pages} {4725} (\bibinfo {year} {2011})}\BibitemShut {NoStop}%
\bibitem [{\citenamefont {Sales}\ \emph {et~al.}(2011)\citenamefont {Sales},
  \citenamefont {Delaire}, \citenamefont {McGuire},\ and\ \citenamefont
  {May}}]{Sales_2011}%
  \BibitemOpen
  \bibfield  {author} {\bibinfo {author} {\bibfnamefont {B.~C.}\ \bibnamefont
  {Sales}}, \bibinfo {author} {\bibfnamefont {O.}~\bibnamefont {Delaire}},
  \bibinfo {author} {\bibfnamefont {M.~A.}\ \bibnamefont {McGuire}}, \ and\
  \bibinfo {author} {\bibfnamefont {A.~F.}\ \bibnamefont {May}},\ }\href
  {\doibase 10.1103/PhysRevB.83.125209} {\bibfield  {journal} {\bibinfo
  {journal} {Phys. Rev. B}\ }\textbf {\bibinfo {volume} {83}},\ \bibinfo
  {pages} {125209} (\bibinfo {year} {2011})}\BibitemShut {NoStop}%
\bibitem [{\citenamefont {Tomczak}\ \emph {et~al.}(2012)\citenamefont
  {Tomczak}, \citenamefont {Haule},\ and\ \citenamefont
  {Kotliar}}]{Tomcsak_2012}%
  \BibitemOpen
  \bibfield  {author} {\bibinfo {author} {\bibfnamefont {J.~M.}\ \bibnamefont
  {Tomczak}}, \bibinfo {author} {\bibfnamefont {K.}~\bibnamefont {Haule}}, \
  and\ \bibinfo {author} {\bibfnamefont {G.}~\bibnamefont {Kotliar}},\
  }\href@noop {} {\bibfield  {journal} {\bibinfo  {journal} {PNAS}\ }\textbf
  {\bibinfo {volume} {109}},\ \bibinfo {pages} {3243} (\bibinfo {year}
  {2012})}\BibitemShut {NoStop}%
\bibitem [{\citenamefont {Takahashi}\ \emph
  {et~al.}(2011{\natexlab{b}})\citenamefont {Takahashi}, \citenamefont {Yasui},
  \citenamefont {Terasaki},\ and\ \citenamefont {Sato}}]{Takahashi_Sato_2011}%
  \BibitemOpen
  \bibfield  {author} {\bibinfo {author} {\bibfnamefont {H.}~\bibnamefont
  {Takahashi}}, \bibinfo {author} {\bibfnamefont {Y.}~\bibnamefont {Yasui}},
  \bibinfo {author} {\bibfnamefont {I.}~\bibnamefont {Terasaki}}, \ and\
  \bibinfo {author} {\bibfnamefont {M.}~\bibnamefont {Sato}},\ }\href@noop {}
  {\bibfield  {journal} {\bibinfo  {journal} {J. Phys. Soc. Jpn.}\ }\textbf
  {\bibinfo {volume} {80}},\ \bibinfo {pages} {054708} (\bibinfo {year}
  {2011}{\natexlab{b}})}\BibitemShut {NoStop}%
\bibitem [{\citenamefont {Wang}\ \emph {et~al.}(2012)\citenamefont {Wang},
  \citenamefont {Hu}, \citenamefont {Warren},\ and\ \citenamefont
  {Petrovic}}]{Wang_2012}%
  \BibitemOpen
  \bibfield  {author} {\bibinfo {author} {\bibfnamefont {K.}~\bibnamefont
  {Wang}}, \bibinfo {author} {\bibfnamefont {R.}~\bibnamefont {Hu}}, \bibinfo
  {author} {\bibfnamefont {J.}~\bibnamefont {Warren}}, \ and\ \bibinfo {author}
  {\bibfnamefont {C.}~\bibnamefont {Petrovic}},\ }\href@noop {} {\bibfield
  {journal} {\bibinfo  {journal} {J. Appl. Phys.}\ }\textbf {\bibinfo {volume}
  {112}},\ \bibinfo {pages} {013703} (\bibinfo {year} {2012})}\BibitemShut
  {NoStop}%
\bibitem [{\citenamefont {Stone}\ \emph {et~al.}(2012)\citenamefont {Stone},
  \citenamefont {Lumsden}, \citenamefont {Nagler}, \citenamefont {Singh},
  \citenamefont {He}, \citenamefont {Sales},\ and\ \citenamefont
  {Mandrus}}]{Stone_2012}%
  \BibitemOpen
  \bibfield  {author} {\bibinfo {author} {\bibfnamefont {M.~B.}\ \bibnamefont
  {Stone}}, \bibinfo {author} {\bibfnamefont {M.~D.}\ \bibnamefont {Lumsden}},
  \bibinfo {author} {\bibfnamefont {S.~E.}\ \bibnamefont {Nagler}}, \bibinfo
  {author} {\bibfnamefont {D.~J.}\ \bibnamefont {Singh}}, \bibinfo {author}
  {\bibfnamefont {J.}~\bibnamefont {He}}, \bibinfo {author} {\bibfnamefont
  {B.~C.}\ \bibnamefont {Sales}}, \ and\ \bibinfo {author} {\bibfnamefont
  {D.}~\bibnamefont {Mandrus}},\ }\href {\doibase
  10.1103/PhysRevLett.108.167202} {\bibfield  {journal} {\bibinfo  {journal}
  {Phys. Rev. Lett.}\ }\textbf {\bibinfo {volume} {108}},\ \bibinfo {pages}
  {167202} (\bibinfo {year} {2012})}\BibitemShut {NoStop}%
\bibitem [{\citenamefont {Sales}\ \emph {et~al.}(2006)\citenamefont {Sales},
  \citenamefont {Jin}, \citenamefont {Mandrus},\ and\ \citenamefont
  {Khalifah}}]{Sales_2006}%
  \BibitemOpen
  \bibfield  {author} {\bibinfo {author} {\bibfnamefont {B.~C.}\ \bibnamefont
  {Sales}}, \bibinfo {author} {\bibfnamefont {R.}~\bibnamefont {Jin}}, \bibinfo
  {author} {\bibfnamefont {D.}~\bibnamefont {Mandrus}}, \ and\ \bibinfo
  {author} {\bibfnamefont {P.}~\bibnamefont {Khalifah}},\ }\href {\doibase
  10.1103/PhysRevB.73.224435} {\bibfield  {journal} {\bibinfo  {journal} {Phys.
  Rev. B}\ }\textbf {\bibinfo {volume} {73}},\ \bibinfo {pages} {224435}
  (\bibinfo {year} {2006})}\BibitemShut {NoStop}%
\bibitem [{\citenamefont {Harada}\ \emph {et~al.}(2004)\citenamefont {Harada},
  \citenamefont {Kanomata}, \citenamefont {Takahashi}, \citenamefont {Nashima},
  \citenamefont {Yoshida},\ and\ \citenamefont {Kaneko}}]{Harada_2004}%
  \BibitemOpen
  \bibfield  {author} {\bibinfo {author} {\bibfnamefont {T.}~\bibnamefont
  {Harada}}, \bibinfo {author} {\bibfnamefont {T.}~\bibnamefont {Kanomata}},
  \bibinfo {author} {\bibfnamefont {Y.}~\bibnamefont {Takahashi}}, \bibinfo
  {author} {\bibfnamefont {O.}~\bibnamefont {Nashima}}, \bibinfo {author}
  {\bibfnamefont {H.}~\bibnamefont {Yoshida}}, \ and\ \bibinfo {author}
  {\bibfnamefont {T.}~\bibnamefont {Kaneko}},\ }\href@noop {} {\bibfield
  {journal} {\bibinfo  {journal} {J. Alloys Compd.}\ }\textbf {\bibinfo
  {volume} {383}},\ \bibinfo {pages} {200} (\bibinfo {year}
  {2004})}\BibitemShut {NoStop}%
\bibitem [{\citenamefont {Neamen}(1992)}]{Neamen_1992}%
  \BibitemOpen
  \bibfield  {author} {\bibinfo {author} {\bibfnamefont {D.~A.}\ \bibnamefont
  {Neamen}},\ }\href@noop {} {\emph {\bibinfo {title} {Semiconductor Physics
  and Devices, Basic Principles}}}\ (\bibinfo  {publisher} {Irwin},\ \bibinfo
  {address} {Boston, MA},\ \bibinfo {year} {1992})\BibitemShut {NoStop}%
\bibitem [{\citenamefont {Yamanouchi}\ \emph {et~al.}(1967)\citenamefont
  {Yamanouchi}, \citenamefont {Mizuguchi},\ and\ \citenamefont
  {Sasaki}}]{Yamanouchi_1967}%
  \BibitemOpen
  \bibfield  {author} {\bibinfo {author} {\bibfnamefont {C.}~\bibnamefont
  {Yamanouchi}}, \bibinfo {author} {\bibfnamefont {K.}~\bibnamefont
  {Mizuguchi}}, \ and\ \bibinfo {author} {\bibfnamefont {W.}~\bibnamefont
  {Sasaki}},\ }\href@noop {} {\bibfield  {journal} {\bibinfo  {journal} {J.
  Phys. Soc. Japan}\ }\textbf {\bibinfo {volume} {22}},\ \bibinfo {pages} {859}
  (\bibinfo {year} {1967})}\BibitemShut {NoStop}%
\bibitem [{\citenamefont {Fritzsche}(1955)}]{Fritzsche_1955}%
  \BibitemOpen
  \bibfield  {author} {\bibinfo {author} {\bibfnamefont {H.}~\bibnamefont
  {Fritzsche}},\ }\href {\doibase 10.1103/PhysRev.99.406} {\bibfield  {journal}
  {\bibinfo  {journal} {Phys. Rev.}\ }\textbf {\bibinfo {volume} {99}},\
  \bibinfo {pages} {406} (\bibinfo {year} {1955})}\BibitemShut {NoStop}%
\bibitem [{\citenamefont {Swartz}(1960)}]{Swartz_1960}%
  \BibitemOpen
  \bibfield  {author} {\bibinfo {author} {\bibfnamefont {G.~A.}\ \bibnamefont
  {Swartz}},\ }\href@noop {} {\bibfield  {journal} {\bibinfo  {journal} {J.
  Phys. Chem. Solids}\ }\textbf {\bibinfo {volume} {12}},\ \bibinfo {pages}
  {245} (\bibinfo {year} {1960})}\BibitemShut {NoStop}%
\bibitem [{\citenamefont {Li}\ \emph {et~al.}(2008)\citenamefont {Li},
  \citenamefont {Qin},\ and\ \citenamefont {Li}}]{Li_2008}%
  \BibitemOpen
  \bibfield  {author} {\bibinfo {author} {\bibfnamefont {H.~J.}\ \bibnamefont
  {Li}}, \bibinfo {author} {\bibfnamefont {X.~Y.}\ \bibnamefont {Qin}}, \ and\
  \bibinfo {author} {\bibfnamefont {D.}~\bibnamefont {Li}},\ }\href@noop {}
  {\bibfield  {journal} {\bibinfo  {journal} {Mater. Sci. Eng.: B}\ }\textbf
  {\bibinfo {volume} {149}},\ \bibinfo {pages} {53} (\bibinfo {year}
  {2008})}\BibitemShut {NoStop}%
\bibitem [{\citenamefont {Li}\ \emph {et~al.}(2009)\citenamefont {Li},
  \citenamefont {Qin}, \citenamefont {Li},\ and\ \citenamefont
  {Xin}}]{Li_2009}%
  \BibitemOpen
  \bibfield  {author} {\bibinfo {author} {\bibfnamefont {H.~J.}\ \bibnamefont
  {Li}}, \bibinfo {author} {\bibfnamefont {X.~Y.}\ \bibnamefont {Qin}},
  \bibinfo {author} {\bibfnamefont {D.}~\bibnamefont {Li}}, \ and\ \bibinfo
  {author} {\bibfnamefont {H.~X.}\ \bibnamefont {Xin}},\ }\href@noop {}
  {\bibfield  {journal} {\bibinfo  {journal} {J. Alloys Compd.}\ }\textbf
  {\bibinfo {volume} {472}},\ \bibinfo {pages} {400} (\bibinfo {year}
  {2009})}\BibitemShut {NoStop}%
\bibitem [{\citenamefont {Sun}\ \emph {et~al.}(2010)\citenamefont {Sun},
  \citenamefont {Oeschler}, \citenamefont {Johnsen}, \citenamefont {Iversen},\
  and\ \citenamefont {Steglich}}]{Sun_Dalton_2010}%
  \BibitemOpen
  \bibfield  {author} {\bibinfo {author} {\bibfnamefont {P.}~\bibnamefont
  {Sun}}, \bibinfo {author} {\bibfnamefont {N.}~\bibnamefont {Oeschler}},
  \bibinfo {author} {\bibfnamefont {S.}~\bibnamefont {Johnsen}}, \bibinfo
  {author} {\bibfnamefont {B.~B.}\ \bibnamefont {Iversen}}, \ and\ \bibinfo
  {author} {\bibfnamefont {F.}~\bibnamefont {Steglich}},\ }\href@noop {}
  {\bibfield  {journal} {\bibinfo  {journal} {Dalton Trans.}\ }\textbf
  {\bibinfo {volume} {39}},\ \bibinfo {pages} {1012} (\bibinfo {year}
  {2010})}\BibitemShut {NoStop}%
\bibitem [{\citenamefont {Ashcroft}\ and\ \citenamefont
  {Mermin}(1976)}]{AshcroftMermin}%
  \BibitemOpen
  \bibfield  {author} {\bibinfo {author} {\bibfnamefont {N.~W.}\ \bibnamefont
  {Ashcroft}}\ and\ \bibinfo {author} {\bibfnamefont {N.~D.}\ \bibnamefont
  {Mermin}},\ }\href@noop {} {\emph {\bibinfo {title} {Solid State Physics}}}\
  (\bibinfo  {publisher} {Thomson Learning Inc},\ \bibinfo {year}
  {1976})\BibitemShut {NoStop}%
\bibitem [{\citenamefont {Kevane}\ \emph {et~al.}(1953)\citenamefont {Kevane},
  \citenamefont {Legvold},\ and\ \citenamefont {Spedding}}]{Kevane_1953}%
  \BibitemOpen
  \bibfield  {author} {\bibinfo {author} {\bibfnamefont {C.~J.}\ \bibnamefont
  {Kevane}}, \bibinfo {author} {\bibfnamefont {S.}~\bibnamefont {Legvold}}, \
  and\ \bibinfo {author} {\bibfnamefont {F.~H.}\ \bibnamefont {Spedding}},\
  }\href {\doibase 10.1103/PhysRev.91.1372} {\bibfield  {journal} {\bibinfo
  {journal} {Phys. Rev.}\ }\textbf {\bibinfo {volume} {91}},\ \bibinfo {pages}
  {1372} (\bibinfo {year} {1953})}\BibitemShut {NoStop}%
\bibitem [{\citenamefont {Paschen}\ \emph {et~al.}(2003)\citenamefont
  {Paschen}, \citenamefont {L{\"u}hmann}, \citenamefont {Langhammer},
  \citenamefont {Trovarelli}, \citenamefont {Wirth}, \citenamefont {Geibel},\
  and\ \citenamefont {Steglich}}]{Paschen_2003}%
  \BibitemOpen
  \bibfield  {author} {\bibinfo {author} {\bibfnamefont {S.}~\bibnamefont
  {Paschen}}, \bibinfo {author} {\bibfnamefont {T.}~\bibnamefont
  {L{\"u}hmann}}, \bibinfo {author} {\bibfnamefont {C.}~\bibnamefont
  {Langhammer}}, \bibinfo {author} {\bibfnamefont {O.}~\bibnamefont
  {Trovarelli}}, \bibinfo {author} {\bibfnamefont {S.}~\bibnamefont {Wirth}},
  \bibinfo {author} {\bibfnamefont {C.}~\bibnamefont {Geibel}}, \ and\ \bibinfo
  {author} {\bibfnamefont {F.}~\bibnamefont {Steglich}},\ }\href@noop {}
  {\bibfield  {journal} {\bibinfo  {journal} {Acta Phys. Pol. B}\ }\textbf
  {\bibinfo {volume} {34}},\ \bibinfo {pages} {359} (\bibinfo {year}
  {2003})}\BibitemShut {NoStop}%
\bibitem [{\citenamefont {Goldsmid}(1986)}]{Goldsmid_1986}%
  \BibitemOpen
  \bibfield  {author} {\bibinfo {author} {\bibfnamefont {H.~J.}\ \bibnamefont
  {Goldsmid}},\ }\href@noop {} {\emph {\bibinfo {title} {Electronic
  Refrigeration}}}\ (\bibinfo  {publisher} {Pion Limited},\ \bibinfo {address}
  {London},\ \bibinfo {year} {1986})\BibitemShut {NoStop}%
\bibitem [{\citenamefont {Sales}\ \emph {et~al.}(2010)\citenamefont {Sales},
  \citenamefont {McGuire}, \citenamefont {Sefat},\ and\ \citenamefont
  {Mandrus}}]{Sales_2010}%
  \BibitemOpen
  \bibfield  {author} {\bibinfo {author} {\bibfnamefont {B.~C.}\ \bibnamefont
  {Sales}}, \bibinfo {author} {\bibfnamefont {M.~A.}\ \bibnamefont {McGuire}},
  \bibinfo {author} {\bibfnamefont {A.~S.}\ \bibnamefont {Sefat}}, \ and\
  \bibinfo {author} {\bibfnamefont {D.}~\bibnamefont {Mandrus}},\ }\href@noop
  {} {\bibfield  {journal} {\bibinfo  {journal} {Physica C}\ }\textbf {\bibinfo
  {volume} {470}},\ \bibinfo {pages} {304} (\bibinfo {year}
  {2010})}\BibitemShut {NoStop}%
\bibitem [{\citenamefont {Hess}\ \emph {et~al.}(2001)\citenamefont {Hess},
  \citenamefont {Baumann}, \citenamefont {Ammerahl}, \citenamefont {B\"uchner},
  \citenamefont {Heidrich-Meisner}, \citenamefont {Brenig},\ and\ \citenamefont
  {Revcolevschi}}]{Hess_2001}%
  \BibitemOpen
  \bibfield  {author} {\bibinfo {author} {\bibfnamefont {C.}~\bibnamefont
  {Hess}}, \bibinfo {author} {\bibfnamefont {C.}~\bibnamefont {Baumann}},
  \bibinfo {author} {\bibfnamefont {U.}~\bibnamefont {Ammerahl}}, \bibinfo
  {author} {\bibfnamefont {B.}~\bibnamefont {B\"uchner}}, \bibinfo {author}
  {\bibfnamefont {F.}~\bibnamefont {Heidrich-Meisner}}, \bibinfo {author}
  {\bibfnamefont {W.}~\bibnamefont {Brenig}}, \ and\ \bibinfo {author}
  {\bibfnamefont {A.}~\bibnamefont {Revcolevschi}},\ }\href {\doibase
  10.1103/PhysRevB.64.184305} {\bibfield  {journal} {\bibinfo  {journal} {Phys.
  Rev. B}\ }\textbf {\bibinfo {volume} {64}},\ \bibinfo {pages} {184305}
  (\bibinfo {year} {2001})}\BibitemShut {NoStop}%
\bibitem [{\citenamefont {Weber}\ and\ \citenamefont
  {Gmelin}(1991)}]{Weber_1991}%
  \BibitemOpen
  \bibfield  {author} {\bibinfo {author} {\bibfnamefont {L.}~\bibnamefont
  {Weber}}\ and\ \bibinfo {author} {\bibfnamefont {E.}~\bibnamefont {Gmelin}},\
  }\href@noop {} {\bibfield  {journal} {\bibinfo  {journal} {Appl. Phys. A}\
  }\textbf {\bibinfo {volume} {53}},\ \bibinfo {pages} {136} (\bibinfo {year}
  {1991})}\BibitemShut {NoStop}%
\bibitem [{\citenamefont {Herring}(1954)}]{Herring_1954}%
  \BibitemOpen
  \bibfield  {author} {\bibinfo {author} {\bibfnamefont {C.}~\bibnamefont
  {Herring}},\ }\href {\doibase 10.1103/PhysRev.96.1163} {\bibfield  {journal}
  {\bibinfo  {journal} {Phys. Rev.}\ }\textbf {\bibinfo {volume} {96}},\
  \bibinfo {pages} {1163} (\bibinfo {year} {1954})}\BibitemShut {NoStop}%
\bibitem [{\citenamefont {Kuroki}\ and\ \citenamefont
  {Arita}(2007)}]{Kuroki_2007}%
  \BibitemOpen
  \bibfield  {author} {\bibinfo {author} {\bibfnamefont {K.}~\bibnamefont
  {Kuroki}}\ and\ \bibinfo {author} {\bibfnamefont {R.}~\bibnamefont {Arita}},\
  }\href@noop {} {\bibfield  {journal} {\bibinfo  {journal} {J. Phys. Soc.
  Jpn.}\ }\textbf {\bibinfo {volume} {76}},\ \bibinfo {pages} {083707}
  (\bibinfo {year} {2007})}\BibitemShut {NoStop}%
\bibitem [{\citenamefont {Goldsmid}\ and\ \citenamefont
  {Sharp}(1999)}]{Sharp_1999}%
  \BibitemOpen
  \bibfield  {author} {\bibinfo {author} {\bibfnamefont {H.~J.}\ \bibnamefont
  {Goldsmid}}\ and\ \bibinfo {author} {\bibfnamefont {J.~W.}\ \bibnamefont
  {Sharp}},\ }\href@noop {} {\bibfield  {journal} {\bibinfo  {journal} {Journal
  of Electronic Materials}\ }\textbf {\bibinfo {volume} {28}},\ \bibinfo
  {pages} {869} (\bibinfo {year} {1999})}\BibitemShut {NoStop}%
\bibitem [{\citenamefont {Blaha}\ \emph {et~al.}(2001)\citenamefont {Blaha},
  \citenamefont {Schwarz}, \citenamefont {Madsen}, \citenamefont {Kvasnicka},\
  and\ \citenamefont {Luitz}}]{wien}%
  \BibitemOpen
  \bibfield  {author} {\bibinfo {author} {\bibfnamefont {P.}~\bibnamefont
  {Blaha}}, \bibinfo {author} {\bibfnamefont {K.}~\bibnamefont {Schwarz}},
  \bibinfo {author} {\bibfnamefont {G.}~\bibnamefont {Madsen}}, \bibinfo
  {author} {\bibfnamefont {D.}~\bibnamefont {Kvasnicka}}, \ and\ \bibinfo
  {author} {\bibfnamefont {J.}~\bibnamefont {Luitz}},\ }\href@noop {}
  {\bibfield  {journal} {\bibinfo  {journal} {WIEN2k, An Augmented Plane Wave +
  Local Orbitals Program for Calculating Crystal Properties (K. Schwarz, Tech.
  Univ. Wien, Austria)}\ } (\bibinfo {year} {2001})}\BibitemShut {NoStop}%
\bibitem [{\citenamefont {Madsen}\ and\ \citenamefont
  {Singh}(2006)}]{Boltztrap}%
  \BibitemOpen
  \bibfield  {author} {\bibinfo {author} {\bibfnamefont {G.~K.~H.}\
  \bibnamefont {Madsen}}\ and\ \bibinfo {author} {\bibfnamefont {D.~J.}\
  \bibnamefont {Singh}},\ }\href@noop {} {\bibfield  {journal} {\bibinfo
  {journal} {Comp. Phys. Comm.}\ }\textbf {\bibinfo {volume} {175}},\ \bibinfo
  {pages} {67} (\bibinfo {year} {2006})}\BibitemShut {NoStop}%
\bibitem [{\citenamefont {Madsen}\ \emph {et~al.}(2003)\citenamefont {Madsen},
  \citenamefont {Schwarz}, \citenamefont {Blaha},\ and\ \citenamefont
  {Singh}}]{Madsen_2003}%
  \BibitemOpen
  \bibfield  {author} {\bibinfo {author} {\bibfnamefont {G.~K.~H.}\
  \bibnamefont {Madsen}}, \bibinfo {author} {\bibfnamefont {K.}~\bibnamefont
  {Schwarz}}, \bibinfo {author} {\bibfnamefont {P.}~\bibnamefont {Blaha}}, \
  and\ \bibinfo {author} {\bibfnamefont {D.~J.}\ \bibnamefont {Singh}},\ }\href
  {\doibase 10.1103/PhysRevB.68.125212} {\bibfield  {journal} {\bibinfo
  {journal} {Phys. Rev. B}\ }\textbf {\bibinfo {volume} {68}},\ \bibinfo
  {pages} {125212} (\bibinfo {year} {2003})}\BibitemShut {NoStop}%
\end{thebibliography}
\end{document}